\documentclass[10pt]{article}

\usepackage{graphicx}    % needed for including graphics e.g. EPS, PS
\usepackage{amsfonts}
\usepackage{amssymb}
\usepackage{graphics}
\usepackage{mathrsfs}
\usepackage{hyperref}
\usepackage{color}

\newtheorem{theorem}{Theorem}[section]

\newtheorem{lemma}{Lemma}[section]

\long\def\symbolfootnote[#1]#2{\begingroup%
\def\thefootnote{\fnsymbol{footnote}}\footnote[#1]{#2}\endgroup} 

\begin{document}

\begin{center}
\Large \textbf{Entrainment of chaos}
\end{center}
{\par\noindent}
\vspace{-0.3cm}
\begin{center}
\normalsize \textbf{M.U. Akhmet$ \symbolfootnote[1]{Corresponding Author Tel.: +90 312 210 5355,  Fax: +90 312 210 2972, E-mail: marat@metu.edu.tr}^{,a}$, M.O. Fen$^a$} \\
\vspace{0.2cm}
\textit{\textbf{\footnotesize$^a$Department of Mathematics, Middle East Technical University, 06800, Ankara, Turkey}}
\vspace{0.1cm}
\end{center}
{\par\noindent}

\begin{center}
\line(1,0){469}
\end{center}

\begin{flushleft}
\textbf{Abstract}
\end{flushleft}

\noindent\ignorespaces
{\par\noindent}
A new phenomenon, entrainment of chaos, which is understood as a seizure of an irregular behavior by limit cycles, is discussed. As a result, chaotic cycles appear if the chaos amplitude is small. Otherwise, the chaos is not necessarily cyclic, if the perturbations are strong and/or diameter of the limit cycle is small. Sensitivity as a main and a unique ingredient is considered and, in addition, period-doubling route to chaos is chosen for extension. The results may be of strong importance for engineering sciences, brainwaves and biomusicology phenomena as well as can be developed for hydrodynamics. Theoretical results are supported by simulations and discussions over Chua's oscillators, entrainment of chaos by toroidal attractors and controlling problems. Moreover, through an example, by means of the Lyapunov functions method, a chaotic attractor is provided. 

\vspace{0.3cm}
\noindent\ignorespaces \textit{\textbf{Keywords:}} Orbital stability, Limit cycle, Chaotic attractor, Sensitivity, Period-doubling cascade, Hopf bifurcation, Toroidal attractor, Chua's oscillator, Chaos control

\begin{center}
\line(1,0){469}
\end{center}

\section{Introduction}

It was Christiaan Huygens who observed that two of his pendulum clocks mounted next to each other on the same support often become synchronized. He called this synchronization tendency as ``entrainment". One can say also about the practice of entraining one's brainwaves to a desired frequency, that is, brainwaves  entrainment \cite{Oster,Walter} or biomusicology as the synchronization of organisms to an external rhythm. Entrainment phenomenon is known also in hydrodynamics  as the movement of one fluid by another. Results of our paper ensure us to say about entrainment of chaos, which can be applied, for example, in economics theory to achieve irregularity in business cycles \cite{Lorenz93}, and to obtain chaotic cycles in electrical circuits by means of Van der Pol equations \cite{Hassard81} and in chemical oscillators such as Belousov-Zhabotinsky reactions in continuous-flow, stirred tank reactors \cite{Field93}. Since, starting with  experiments of Huygens, the phenomena is mainly related to cyclic motions, more precisely, entrainment of chaos by limit cycles is under discussion in our paper. Thus, we demonstrate that the idea of entrainment is not concerning only frequency, period or phase \cite{Pik01,Nadal09}, but chaos, also. In our study the concept of entrainment is less symmetric than that for Huygens synchronization. It is, rather similar to the phenomenon in biomusicology or hydrodynamics. Nevertheless, unidirectional couplings have been intensively considered in physical studies \cite{Nadal09}-\cite{Keller02}. We hope that our investigations will  be useful first of all for neural sciences, then they are of some interest for chemistry, mechanics, electronics and population dynamics. Shortly speaking, they can be applied and developed for any research, where limit cycles have been observed.  
 
In general, we consider systems, which admit well arranged, steady motions and subject to them  external irregular perturbations. The question is, how their motions are tuned to the external influences? If the external forces are regular, then reactions are well analyzed in classical literature. For periodic perturbations - periodic responses, if perturbations are almost periodic, then responses are almost periodic, etc \cite{Hale63,Corduneanu09}. What about chaotic perturbations? An answer to this question has been introduced in our papers \cite{Akh8}-\cite{Akh11}, where we discuss the problem if systems are with asymptotically stable equilibriums, and found that solutions admit the same type of chaos as perturbations do.

The paper \cite{Akh8}, where we discuss the mechanism of morphogenesis of chaos, informs us about replication of specific types of chaos, such as Devaney and Li-Yorke chaos and chaos obtained through period-doubling cascade, in continuous-time systems with arbitrary high dimensions. In this process, we take into account the generator-replicator systems such that the generator is considered as a system of the form
\begin{eqnarray} \label{generator}
x'=F(t,x),
\end{eqnarray}
where $F: \mathbb R \times \mathbb R^m \to \mathbb R^m $ is a continuous function in all its arguments 
and the replicator is assumed to have the form
\begin{eqnarray} \label{replicator}
y'=Ay+g(x,y),
\end{eqnarray}
where $g: \mathbb R^m \times \mathbb R^n \to \mathbb R^n$ is a continuous function in all its arguments and the constant $n\times n $ real valued matrix $A$ has real parts of eigenvalues all negative. 

The rigorous results of the morphogenesis (extension) mechanism emphasize that  system $(\ref{replicator})$ is chaotic in the same way as system $(\ref{generator}).$ Replication of chaos through intermittency is also demonstrated through simulations in paper \cite{Akh8}, where one can find new definitions for chaotic sets of functions, and precise descriptions for the ingredients of Devaney and Li-Yorke chaos in continuous-time dynamics, which are used as tools for the study of morphogenesis of chaos.

%However, in the present article, we consider systems with orbitally stable limit cycles and perturb them with chaos, and as a result we achieve chaotic behavior around limit cycles, providing a cyclical character. We apply the technique of Andronov-Witt Theorem on orbital stability of cycles \cite{Farkas2010}, and this is a different and a more difficult task, than to study chaos near fixed points. By elaborating the proof of the theorem, we give more details to behavior of  motions near the limit cycle, and then get benefits of this.  
 
Distinctively from the morphogenesis mechanism where we considered chaos near fixed points, in the present article, we take into account systems with orbitally stable limit cycles, and perturb them with chaos. As a result we obtain \textit{chaotic cycles}, that is, motions which behave cyclically and chaotically simultaneously. Since one achieves appearance of chaotic behavior near limit cycles, the sizes of the attained chaotic attractors are possibly larger than the ones considered in paper \cite{Akh8}. Even one can say that the chaos extension procedure developed in the present article is global, while it is local in the former one. 

The theoretical novelty of our analysis lies in the fact that we apply the complex technique of the proof of Andronov-Witt theorem \cite{Farkas2010} to indicate the boundedness of solutions, and describe in detail the cyclical behavior of the trajectories of the chaotically forced system in a neighborhood of the limit cycle. Extension of sensitivity and chaos through period-doubling cascade are also elaborated rigorously in the paper.

Entrainment of chaos should be understood  as the seize of chaos by   the vector field  near the limit  cycle. A novelty of the entrainment process, which is important for applications, is the combination of a chaotic behavior with a cyclical one. Cyclical behavior of the motions can disappear if the applied perturbations are strong and/or diameter of the limit cycle is small. This type of behavior is discussed in the third section of the paper. We prove presence of chaos through indicating the  period-doubling cascade, and sensitivity,  and recognize that observation of other ingredients of chaos, transitivity, density  of periodic motions occurs as a more difficult task, than we  presupposed. Nevertheless, it is known that  \cite{Rob95, Lorenz63} sensitivity is the  main ingredient of chaos, since it  assumes unpredictability. We leave the discussion for the attractiveness of  chaotic solutions, but we consider the problem in an example. 
 
It is very important to comprehend how one can develop present results in further investigations. From this point of view, in the sixth section of our paper, we place simulations of chaotically perturbed toroidal attractor, which results as a chaotic torus, but we do not give a rigorous mathematical basis for this phenomena. Moreover, we perform our method to perturb chaotically stable Chua's oscillators to attain a new chaotic Chua's attractor. Finally, to observe unstable periodic solutions present in the generated chaos, we give an example for the application of Pyragas control method.

To present the main idea of the paper, we continue with an example of an oscillating chemical reaction as a possible model for our theory. Paper \cite{Lengyel90} considers the chlorine dioxide-iodine-malonic acid $(ClO_2-I_2-MA)$ chemical reaction which possesses the following three component reactions:
\begin{eqnarray}
\begin{array}{l} \label{reactions}
MA+I_2  \longrightarrow  IMA+I^- + H^+, \\
ClO_2+I^-  \longrightarrow  ClO_2^- + \frac{1}{2} I_2, \\
ClO_2^-+4I^-+4H^+  \longrightarrow  Cl^- + 2I_2+2H_2O.
\end{array}
\end{eqnarray}

The rate equations corresponding to reactions $(\ref{reactions})$ are given by
\begin{eqnarray}
\begin{array}{l} \label{rate_eqns}
\displaystyle \frac{d[I_2]}{dt} = -\frac{ k_{1a} [MA][I_2]}{k_{1b}+[I_2]},  \\
\displaystyle \frac{d[ClO_2]}{dt}=-k_2[ClO_2][I^-], \\
\displaystyle \frac{d[ClO_2^-]}{dt}=-k_{3a}[ClO_2^-][I^-][H^+] -\displaystyle \frac{k_{3b}[ClO_2^-][I_2][I^-]}{\varsigma+[I^-]^2},
\end{array}
\end{eqnarray}
where $k_{1a},k_{1b},k_{2},k_{3a},k_{3b}$ are the rate constants and $\varsigma^{1/2}$ represents the level of $[I^-]$ above which the inhibitory effect of iodine ion becomes significant.

Due to the complicatedness for analytic calculations, after approximating the concentrations of slow reactants $MA, I_2$ and $ClO_2$ as constants and making reasonable simplifications and nondimensionalizations, Lengyel, R$\acute{a}$bai and Epstein \cite{Lengyel90} reduced  system $(\ref{rate_eqns})$ to the $2-$dimensional system
\begin{eqnarray}
\begin{array}{l} \label{chem_model1}
u_1'=a-u_1-\displaystyle\frac{4u_1u_2}{1+u_1^2}, \\
u_2'=bu_1\left(1-\displaystyle\frac{u_2}{1+u_1^2}\right),
\end{array}
\end{eqnarray}
where $u_1$ and $u_2$ represent the dimensionless concentrations of $I^-$ and $ClO_2^-$ ions, and the parameters $a>0$ and $b>0$ depend on the empirical rate constants and on the concentrations assumed for the slow reactants.  

It can be verified that for a given value of the parameter $a,$ system $(\ref{chem_model1})$ undergoes a Hopf bifurcation at the parameter value $b=b_0\equiv3a/5-25/a$ such that when $b>b_0,$ all trajectories spiral into the stable fixed point $(u_1^*,u_2^*)=(a/5,1+a^2/25),$ while for $b<b_0$ trajectories are attracted to an orbitally stable limit cycle. In that case, if we consider system $(\ref{chem_model1})$ with the coefficient $a=11,$ Hopf bifurcation occurs for $b_0=238/55$ and  consequently, for $b=2.1,$ an orbitally stable limit cycle takes place \cite{Strogatz94}.

Next, we take into account the Birkhoff-Shaw chaotic attractor \cite{Th02,Shaw81} which is generated by the system of differential equations
\begin{eqnarray}
\begin{array}{l} \label{chem_model2}
x_1'=0.7x_2+10x_1(0.1-x_2^2), \\
x_2'=-x_1+0.25 \displaystyle \sin(1.57t). 
\end{array}
\end{eqnarray}

We note that system $(\ref{chem_model2})$ is in analogy with Van der Pol type equations such that if the periodic forcing term $0.25 \displaystyle \sin(1.57t)$ is removed from the second line of $(\ref{chem_model2}),$ then under the transformation $x=\sqrt{10}x_2, x'=-\sqrt{10}x_1,$ one attains the Van der Pol equation
\begin{eqnarray} \label{vanderpol}
x''+\alpha_0(x^2-1)x'+w_0^2x=A_0\displaystyle \sin(k_0 t),
\end{eqnarray}
with the coefficients $\alpha_0=1,w_0=\sqrt{0.7}$ and $A_0=0.$ The periodic forcing is used for the velocity term in system $(\ref{chem_model2}),$ while the acceleration is periodically driven in Van der Pol oscillators in the form of equation $(\ref{vanderpol})$. This type of forcing is unusual in mechanical systems, but can be observed in electrical or chemical systems \cite{Th02}.

Making use of system $(\ref{chem_model2})$ as the source of chaos and combining with system $(\ref{chem_model1})$ in a unidirectional way, we set up the following $4-$dimensional system
\begin{eqnarray}
\begin{array}{l} \label{chem_model3}
x_1'=0.7x_2+10x_1(0.1-x_2^2), \\
x_2'=-x_1+0.25 \displaystyle \sin(1.57t), \\
x_3'=11-x_3-\displaystyle\frac{4x_3x_4}{1+x_3^2}+1.2\displaystyle \tan \left(\displaystyle\frac{x_1}{2}\right), \\
x_4'=2.1 x_3\left(1-\displaystyle\frac{x_4}{1+x_3^2}\right)+0.8x_2.
\end{array}
\end{eqnarray}

Practically, it may not be possible to obtain a chemical reaction which admits the rate equations such as in system $(\ref{chem_model3}).$ However, in this exemplification, our purpose is to realize the feasibility of extension of irregular behavior in system  $(\ref{chem_model1})$ from an arbitrary source of chaos or \textit{chaotic functions}, in the case of a unidirectional perturbation is actualized as in system $(\ref{chem_model3}).$
 
The rigorous results of the present paper indicate that system $(\ref{chem_model3})$ possesses chaotic motions in the $4-$dimensional phase space  and the projection of the chaotic attractor  of system $(\ref{chem_model3})$ on the $x_3-x_4$ plane takes place  around the orbitally stable limit cycle of system $(\ref{chem_model1})$ with the specified coefficients. The appearance of chaotic behavior around the limit cycle is an indicator of entrainment of chaos.
 
\begin{figure}[ht] 
\centering
\includegraphics[width=10.5cm]{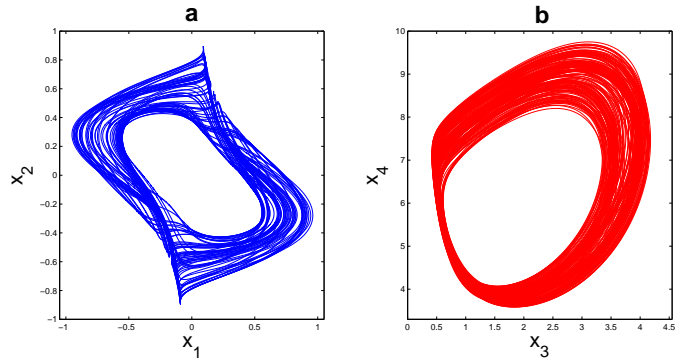}
\caption{The $2-$dimensional projections on the $x_1-x_2$ and $x_3-x_4$ planes of the chaotic attractor of system $(\ref{chem_model3})$ are presented in pictures (a) and (b), respectively. The picture in $(a)$ is, in fact, the Birkhoff-Shaw chaotic attractor produced by system $(\ref{chem_model2}),$ while Figure $\ref{shaw1}, (b)$ represents the chaotic attractor generated around the orbitally stable limit cycle of system $(\ref{chem_model1}),$ where $a=11$ and $b=2.1.$ The irregular structure observed around the limit cycle is a manifestation of entrainment of chaos.}
\label{shaw1}
\end{figure}

\begin{figure}[ht] 
\centering
\includegraphics[width=12.5cm]{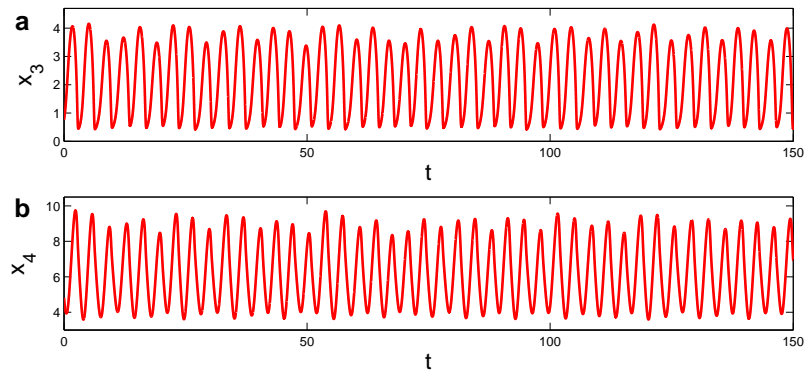}
\caption{The graphs of the $x_3$ and $x_4$ coordinates of system $(\ref{chem_model3})$ are presented. The pictures in $(a)$ and $(b)$ both present the irregular behavior of the coordinates.}
\label{shaw2}
\end{figure}

To provide an illustration for the mentioned behavior of system $(\ref{chem_model3}),$ in Figure $\ref{shaw1},$ the $2-$dimensional projections of the trajectory of system $(\ref{chem_model3})$ with the initial data $x_1(0)=0.2,  x_2(0)=0.3,  x_3(0)=0.75, x_4(0)=4.82$ are pictured. Figure $\ref{shaw1}, (a)$ shows the projection on the $x_1-x_2$ plane and this picture  represents, in fact, the Birkhoff-Shaw chaotic attractor produced by system $(\ref{chem_model2}).$ On the other hand, Figure $\ref{shaw1}, (b)$  reveals the process of entrainment of chaos such that  the chaotic attractor generated around the orbitally stable limit cycle of system $(\ref{chem_model1}),$ where $a=11$ and $b=2.1,$ is illustrated in the picture. Moreover, the irregular behavior in the $x_3$ and $x_4$ coordinates of system $(\ref{chem_model3})$ generated in process of time is pictured in Figure $\ref{shaw2}.$ 

\section{Preliminaries} \label{prelim}

Throughout the paper $\mathbb R$ and $\mathbb R_+$ will denote the set of real numbers and the interval $[0,\infty),$ respectively.

Let us consider the system  
\begin{eqnarray}\label{1}
x'=F(t,x),
\end{eqnarray}
where $F:\mathbb R_+ \times \mathbb R^{m}\to \mathbb R^{m}$ is a continuous function in all its arguments, and take into account the system
\begin{eqnarray} 
\begin{array}{l}\label{3}
u'=f(u)
\end{array}
\end{eqnarray}
where $f:\mathbb R^{n} \to \mathbb R^{n}$ is a continuously differentiable function.

To adapt system $(\ref{3})$  in the entrainment process,  we perturb this system through the solutions of $(\ref{1}),$  and finally obtain the system of the form
\begin{eqnarray}\label{2}
y'=f(y)+\mu g(x),
\end{eqnarray}
where $\mu$ is a nonzero real number  and $g:\mathbb R^{m} \to \mathbb R^{n}$ is a continuous function.

We mainly assume that system $(\ref{1})$ admits a chaotic attractor which is understood as a collection of functions such that each element of the set is called a \textit{chaotic function}. More  detailed description of  chaotic functions is provided in \cite{Akh8}. In this case, there exists a positive real number $M$ such that $\sup_{t \in \mathbb R_+} \left\|x(t)\right\| \leq M,$ for each chaotic solution $x(t)$ of system $(\ref{1}).$ The chaotic functions may be irregular as well as regular (periodic) such that skeleton of the attractor consists of infinitely many {\it unstable}  periodic solutions. This phenomena is true for both Devaney's concept of chaos \cite{Dev90,Palmer00} as well as for the chaos obtained through period-doubling cascade.

Our another assumption is that the non-linear autonomous system $(\ref{3})$ possesses a non-constant $T-$periodic solution $p(t),$ for some positive real number $T$, and we consider  system $(\ref{2})$ in a neighborhood of the orbit
\begin{eqnarray} 
\begin{array}{l}\label{cycle}
\gamma=\left\{u\in \mathbb R^n: u=p(t), t\in [0,T] \right\}.
\end{array}
\end{eqnarray}

It is clear  that $p'(t)$ is a non-trivial $T-$periodic solution of the variational system 
\begin{eqnarray} 
\begin{array}{l}\label{4}
v'=A(t)v,
\end{array}
\end{eqnarray}
where $A(t)=f'(p(t))$ is an $n \times n$ real, continuous, $T-$periodic matrix function, and consequently the number $1$ is a characteristic multiplier of system $(\ref{4}).$

In what follows, we assume that the number $1$ is a simple characteristic multiplier of the variational system $(\ref{4})$ and the remaining $n-1$ characteristic multipliers are in modulus less than one. In this case, by Andronov-Witt Theorem \cite{Farkas2010}, the periodic solution $p(t)$ of system $(\ref{3})$ is asymptotically orbitally stable, having the asymptotic phase property.

The following conditions will be used throughout the paper.
\begin{enumerate}
\item[\bf (A1)] There exists a positive real number $L_f$ such that 
$
\left\|f(y_1)-f(y_2)\right\| \leq L_f\left\|y_1-y_2\right\|,
$ 
for all $y_1,y_2 \in \mathbb R^{n};$ 
\item[\bf (A2)] There exists a positive real number $L_g$ such that 
$ 
\left\|g(x_1)-g(x_2)\right\| \geq L_g\left\|x_1-x_2\right\|,
$
for all $x_1,x_2 \in \mathbb R^{m};$ 
\item[\bf (A3)] There exist positive real numbers $M_F, M_g$ such that 
$
\displaystyle \sup_{t\in \mathbb R_+, x\in \mathbb R^m} \left\|F(t,x)\right\|\le M_F
$
and 
$
\displaystyle \sup_{x\in \mathbb R^m} \left\|g(x)\right\|\le M_g.
$
\end{enumerate}

We emphasize that for an arbitrary solution $u(t)$ of system $(\ref{3}),$  the function $z(t)=u(t)-p(t)$ is a solution of the system 
\begin{eqnarray} 
\begin{array}{l}\label{5}
z'=A(t)z + \varphi(t,z),
\end{array}
\end{eqnarray}
where $\varphi(t,z)=f(p(t)+z)-f(p(t))-f'(p(t))z.$ It is clear that $\varphi(t+T,z)=\varphi(t,z)$ and $\varphi (t,0)=\varphi_z (t,0)=0,$ for all $t \in \mathbb R_+,$ and continuity of $f'$ implies that $\varphi_z(t,z)=o(1)$ as $z \to 0$ uniformly in $t \in \mathbb R_+.$

Similarly, for any given solution $x(t)$ of system $(\ref{1}),$ if $\phi(t)$ is an arbitrary solution of the equation $y'=f(y)+\mu g(x(t)),$ then $w(t)=\phi(t)-p(t)$ is a solution of the following system 
\begin{eqnarray} 
\begin{array}{l}\label{6}
w'=A(t)w + \varphi(t,w) + \mu g(x(t)).
\end{array}
\end{eqnarray}

Since we take advantage of  the proof of Andronov-Witt Theorem, we continue with a brief explanation of the technique  used in \cite{Farkas2010}.

According to our assumption that system $(\ref{4})$ admits the number $1$ as a simple characteristic multiplier and the remaining $n-1$ characteristic multipliers are in modulus less than one, system $(\ref{4})$ has a real fundamental matrix $\Phi(t)$ of the form 
\begin{eqnarray} 
\Phi(t)=P(t) \left(
\begin {array}{ccc}  
1 & 0 \label{7}\\
\noalign{\medskip}
0 & e^{B_1t}
\end {array}
\right),
\end{eqnarray}
where $P(t)$ is a regular, continuously differentiable $T-$periodic matrix and $B_1$ is an $(n-1) \times (n-1)$ matrix whose eigenvalues are all with negative real parts.

In the rest of the paper, without loss of generality, we suppose that the coordinates have been translated and rotated such that the periodic solution $p(t)$ of system $(\ref{3})$ satisfies that $p(0)=0$ and $p'(0)=\left(\bar{p}_1,0,0,\ldots,0 \right)$ for some positive real number $\bar{p}_1.$

By the help of the fundamental matrix introduced by equation $(\ref{7}),$ we define the function
\begin{eqnarray}
 G(t,s)=\left\{\begin{array}{ll} \Phi(t)\left(
\begin {array}{ccc}  
0 & 0 \label{8}\\
\noalign{\medskip}
0 & I_{n-1}
\end {array}
\right)\Phi^{-1}(s), & ~t>s \\
-\Phi(t)\left(
\begin {array}{ccc}  
1 & 0 \\
\noalign{\medskip}
0 & O_{n-1}
\end {array}
\right)\Phi^{-1}(s), & ~t<s, 
\end{array} \right.
\end{eqnarray} 
where $I_{n-1}$ and $O_{n-1}$ denote the $(n-1)\times(n-1)$ identity and the zero matrices, respectively.

Now, let us choose a real number $\alpha>0$ such that $Re(\alpha_k)<-\alpha,$ for all $k=1,2,\ldots,n-1,$ where $\alpha_1,\alpha_2,\ldots,\alpha_{n-1}$ denote the eigenvalues of the matrix $B_1.$
In this case, there exists a real number $K>0$ such that $\left\|G(t,s)\right\| \leq K e^{-\alpha(t-s)}$ for $t>s$ and $\left\|G(t,s)\right\| \leq K$ for $t \leq s.$ 

Let $L_{\varphi}=\frac{\alpha}{8K}.$ Since $\varphi_z(t,z)=o(1)$ as $z \to 0$ uniformly in $t \in \mathbb R_+,$ there exists a number $\widetilde{\delta}(L_{\varphi})>0$ such that if $\left\|z_1\right\|< \widetilde{\delta}(L_{\varphi}),$ $\left\|z_2\right\|< \widetilde{\delta}(L_{\varphi}),$ then the inequality $\left\|\varphi(t,z_1)-\varphi(t,z_2)\right\|\leq L_{\varphi} \left\|z_1-z_2\right\|$ is satisfied uniformly in $t \in \mathbb R_+.$

Suppose that $a=\left( 0,a_2,a_3,\ldots,a_n \right)$ is an $n-$dimensional vector which is orthogonal to $p'(0).$ It follows from equation $(\ref{7})$ that there exists a positive real number $K_1$ such that $\left\|\Phi(t)a\right\|\leq K_1\left\|a\right\|e^{-\alpha t},$ for all $t\in \mathbb R_+.$ Using the method presented in \cite{Farkas2010}, one can verify that if $\left\| a \right\| <\displaystyle \frac{\widetilde{\delta}(L_{\varphi})}{2K_1},$ then a solution $z(t,a)$ of system $(\ref{5})$
exists on $\mathbb R_+$ and satisfies the relation
\begin{eqnarray} 
\begin{array}{l}\label{10}
z(t,a)=\Phi(t)a+\displaystyle \int_0^{\infty} G(t,s) \varphi(s,z(s,a))ds.
\end{array}
\end{eqnarray}
Moreover, the solution $z(t,a)$ satisfies the inequality
\begin{eqnarray} 
\begin{array}{l}\label{11}
\left\|z(t,a)\right\| \leq 2K_1\left\|a\right\|e^{-\alpha t /2},
\end{array}
\end{eqnarray}
which is valid for $t \in \mathbb R_+.$

In the sequel, let us denote by $\zeta(t,\zeta_0)$ the solution of system $(\ref{3})$ which corresponds to the initial condition $\zeta(0,\zeta_0)=\zeta_0.$

For any given initial data $\zeta_0,$ the solution $\zeta(t,\zeta_0)$ provides the relation $\zeta(t,\zeta_0)=z(t,a)+p(t),$ where $z(t,a)$ is a solution of system $(\ref{5})$ with the initial condition $z(0,a)=\zeta_0.$

According to equation $(\ref{10}),$ we have
\begin{eqnarray} \label{initial_vec}
\begin{array}{l}
\zeta_0=  \Phi(0)a+\displaystyle \int_{0}^{\infty} G(0,s) \varphi(s,z(s,a)) ds  = P(0)a-\widetilde{h}(a),
\end{array}
\end{eqnarray}
where
\begin{eqnarray} \label{function_h_tilde}
\widetilde{h}(a)=P(0) \left(
\begin {array}{ccc}  
1 & 0 \\
\noalign{\medskip}
0 & O_{n-1}
\end {array}
\right) \displaystyle \int_{0}^{\infty} P^{-1}(s) \varphi(s,z(s,a)) ds, 
\end{eqnarray} 
provided that the inequality $\left\| a \right\| < \displaystyle \frac{\widetilde{\delta}(L_{\varphi})}{2K_1}$ is valid. Since the first column of the matrix $P(0)$ is $[1,0,\ldots,0],$ it can be verified that the function $\widetilde{h}(a)$ can be written in the form
\begin{eqnarray*}
\widetilde{h}(a)=\left(\widetilde{h}_1(a_2,\ldots,a_n),0,\ldots,0\right),
\end{eqnarray*}
for some continuously differentiable function $\widetilde{h}_1.$ It is easy to show that $\widetilde{h}(a)=o(\left\|a\right\|)$ as $a$ tends to zero. Suppose that  $\zeta_0=(\zeta_1^0,\zeta_2^0,\ldots,\zeta_n^0)$ and $p_{ij}$ are the coordinates of the matrix $P(0),$ where $i,j=1,2,\ldots,n.$ In this case equation $(\ref{initial_vec})$ is equivalent to the following system of $n$ equations
\begin{eqnarray} \label{initial_vec2}
\begin{array}{l}
\zeta_1^0=\displaystyle \sum_{j=2}^{n} p_{1j}a_j-\widetilde{h}_1(a_2,\ldots,a_n),\\
\zeta_i^0=\displaystyle \sum_{j=2}^{n} p_{ij}a_j, ~i=2,3,\ldots,n.
\end{array}
\end{eqnarray}
The last $(n-1)$ equations presented in $(\ref{initial_vec2})$ form a linear, regular, one-to-one mapping between the last $n-1$ coordinates of the vectors $a$ and $\zeta_0.$ Expressing $a_2,\ldots,a_n$ in terms of $\zeta_2^0,\ldots,\zeta_n^0$ and substituting these values into the first equation presented in $(\ref{initial_vec2}),$ one can obtain an equation of the form 
\begin{eqnarray}
\zeta_1^0+\displaystyle \sum_{i=2}^n q_i \zeta_i^0 - h(\eta_2^0,\zeta_3^0,\ldots,\zeta_n^0)=0,
\end{eqnarray} 
where $q_i,$ $i=2,\ldots,n,$ are constants and $h$ is a continuously differentiable function such that 
\begin{displaymath}
h(\zeta_2^0,\ldots,\zeta_n^0)=o\left( \left( \sum_{i=2}^n (\zeta_i^0)^2  \right)^{1/2} \right).
\end{displaymath}

Suppose that $S \subset \mathbb R^n$ is an $(n-1)$ dimensional, $C^1$ manifold determined by the equation
\begin{eqnarray} 
\begin{array}{l}\label{12}
x_1+\displaystyle \sum_{i=2}^{n} q_ix_i-h(x_2,x_3,\ldots,x_n)=0.
\end{array}
\end{eqnarray}
We note that the manifold $S$ is a hypersurface in a neighborhood of the origin, crossing the orbit $\gamma,$ which is  defined by equation
$(\ref{cycle}),$ transversally such that for any solution $\zeta(t,\zeta_0)$ of system $(\ref{3})$ with $\zeta_0\in S,$ we have $\left\|\zeta(t,\zeta_0)-p(t)\right\|$ tends to zero exponentially as $t$ tends to infinity.

Let $\overline{\epsilon}=\displaystyle \frac{1}{2\left\|P^{-1}(0)\right\|}.$ Since the function $\widetilde{h}(a)$ defined by equation $(\ref{function_h_tilde})$ satisfies the property $\widetilde{h}(a)=o(\left\|a\right\|)$ as $a$ tends to zero, it is possible in this case to find a real number $\overline{\delta} \left(\overline{\epsilon}\right)>0$ such that if $\left\|a\right\| < \displaystyle \min \left\{ \widetilde{\delta}(L_{\varphi})/(2K_1), \overline{\delta} \left(\overline{\epsilon}\right)   \right\},$ then the inequality
\begin{eqnarray} \label{ineq1}
\left\|\widetilde{h}(a)\right\| < \overline{\epsilon} \left\|a\right\| 
\end{eqnarray} 
is valid. In the remaining parts of the paper, we suppose that $\left\|a\right\| < \displaystyle \min \left\{ \frac{\widetilde{\delta}(L_{\varphi})}{2K_1}, \overline{\delta} \left(\overline{\epsilon}\right)   \right\}.$

The next section is devoted for the existence and behavior of bounded solutions of system $(\ref{2}).$

\section{Entrainment and boundedness of solutions} \label{sec_boundedness}

In Lemma $(\ref{lemma1}),$ we detail the behavior of trajectories  near the limit cycle,  whose orbital stability is ensured by Andronov-Witt Theorem, and then show that the perturbed system admits a set of bounded solutions near the limit cycle through Theorem $\ref{boundedness}.$ Thus, we shall prepare a carrier for chaotic solutions.

\begin{lemma} \label{lemma1}
For each real number $l\in (0,1),$ there exists a natural number $n_0=n_0(l)$ such that for any solution $\zeta(t,\zeta_0)$ of system $(\ref{3})$ with $\zeta_0 \in S,$ the inequality $\left\|\zeta(n_0T,\zeta_0)\right\|\leq l\left\|\zeta_0\right\|$ is satisfied. 
\end{lemma}

\noindent \textbf{Proof.}
Let us fix a solution $\zeta(t,\zeta_0)$ of system $(\ref{3})$ such that the initial vector $\zeta_0$ belongs to the surface $S.$ Making use of the equation $(\ref{initial_vec})$ and the inequality $(\ref{ineq1})$ one can attain that 
\begin{eqnarray*}
&& \left\| \zeta_0 \right\| \geq \left\|P(0)a\right\| - \left\| \widetilde{h}(a) \right\| \\
&& \geq \frac{\left\|a\right\|}{\left\|P^{-1}(0)\right\|} - \overline{\epsilon} \left\|a\right\| \\
&& = \frac{\left\|a\right\|}{2\left\|P^{-1}(0)\right\|}.
\end{eqnarray*}

The last inequality implies that $\left\|a\right\| \leq 2 \left\|P^{-1}(0)\right\| \left\| \zeta_0 \right\|.$ In this case, by means of $(\ref{11}),$ it can be verified that
\begin{eqnarray} \label{14}
\begin{array}{l}
 \left\|\zeta(t,\zeta_0)-p(t)\right\| = \left\|z(t,a) \right\|  \leq 2K_1 \left\|a\right\| e^{-\alpha t/2}  \leq  4K_1\left\|P^{-1}(0)\right\|   \left\| \zeta_0 \right\|   e^{-\alpha t/2},
\end{array}
\end{eqnarray}
for all $t\in \mathbb R_+.$

Now, let us fix an arbitrary number $l \in (0,1).$ It is possible to find a natural number $n_0=n_0(l)$ such that the inequality $4K_1\left\|P^{-1}(0)\right\|e^{-\alpha T n_0/2}<l$ holds.
Making use of $(\ref{14})$ we obtain that 
\begin{eqnarray*}
 \left\|\zeta(n_0T,\zeta_0)-p(n_0T)\right\|  \leq  4K_1\left\|P^{-1}(0)\right\| \left\| \zeta_0 \right\| e^{-\alpha T n_0/2}   < l \left\| \zeta_0 \right\|.
\end{eqnarray*}
Since $p(n_0T)=0,$ we have $\left\|\zeta(n_0T,\zeta_0)\right\|<l \left\|\zeta_0\right\|.$ 

The lemma is proved. $\square$

In the rest of the paper, for a given solution $x(t)$ of system $(\ref{1}),$ the function $\eta_{x(t)}(t,\eta_0)$ will stand for the solution of system $y'=f(y)+\mu g(x(t))$ satisfying the initial condition $\eta_{x(t)}(0,\eta_0)=\eta_0.$ Furthermore, $B_r$ will denote the open ball centered at the origin with radius $r>0.$ 

The next theorem signifies not only a boundedness criteria for solutions of system $(\ref{2}),$  but also  their  cyclical behavior.

\begin{theorem} \label{boundedness}
There exist a nonzero real number $\mu,$ a positive real number $\delta$ and a natural number $n_0$ such that any solution $\eta_{x(t)}(t,\eta_0)$ of system $(\ref{2}),$ where $x(t)$ is a chaotic solution of system $(\ref{1})$ and $\eta_0\in B_{\delta},$ is bounded on $\mathbb R_+$ such that for all $t\in \mathbb R_+$ the inequality
\begin{eqnarray} \label{lemma_formula}
\begin{array}{l}
\left\|\eta_{x(t)}(t,\eta_0)-p(t)\right\| \leq \displaystyle \frac{\left|\mu\right|M_g}{L_f} \left(e^{L_f (n_0+2)T}-1\right) + H(\delta,\rho), 
\end{array}
\end{eqnarray}
holds, where $\rho=\displaystyle \max_{t \in [0,T]} \left\|p(t)\right\|,$ and $ H(\delta, \rho) = \displaystyle \max (\delta+\rho) \left\{e^{2L_f T}, 4K_1\left\|P^{-1}(0)\right\|  \left[\rho + \left(\delta + \rho \right) e^{2L_f T}\right]\right\}.$
\end{theorem}

\noindent \textbf{Proof.}
The essence of the proof is to determine the numbers $\mu \neq 0$ and $\delta >0$ such that for any given solution $\eta_{x(t)}(t,\eta_0),$ where $x(t)$ is a chaotic solution of system $(\ref{1})$ and $\eta_0\in B_{\delta},$ there exists a sequence $\left\{\theta_i\right\},$ $\theta_i\to \infty$ as $i \to \infty,$ such that $\eta_{x(t)}(t,\eta_0) $ enters that ball infinitely many times at the moments $t=\theta_i,$ and consequently the collection of all such functions are uniformly bounded. For our purpose, we will make use of the solutions of system $(\ref{3})$ which have the same initial data with the chosen function $\eta_{x(t)}(t,\eta_0)$ at the moments $\theta_i.$ We note that the sequence $\left\{\theta_i\right\},$ which will be constructed in the proof, depends on the function $\eta_{x(t)}(t,\eta_0),$ that is, depends on both $x(t)$ and the initial data $\eta_0 \in B_{\delta}.$

Since the orbit $\gamma$ of the periodic solution $p(t)$ of system $(\ref{3})$ intersects the surface $S$ transversally, there exists a real number $\epsilon_1>0$ such that if $\left\|\zeta(t,\zeta_0)-p(t)\right\|<\epsilon_1$ for each $t \in [0,2T],$ then $\zeta(t,\zeta_0)$ intersects $S$ at some moment $t_1\in [0,2T].$ Suppose that a positive number $\delta=\delta(\epsilon_1)$ is chosen such that $\delta \leq \displaystyle \epsilon_1 e^{-2L_f T}$ and let an arbitrary $\zeta_0 \in B_{\delta}$ be given.

The solution $\zeta(t,\zeta_0)$ and the periodic solution $p(t)=\zeta(t,0)$ of system $(\ref{3})$ satisfy the relations
\begin{eqnarray*}
\zeta(t,\zeta_0)=\zeta_0 + \displaystyle \int_{0}^{t} f(\zeta(s,\zeta_0)) ds 
\end{eqnarray*}
and
\begin{eqnarray*}
p(t)=\displaystyle \int_{0}^{t} f(p(s)) ds, 
\end{eqnarray*}
respectively. Therefore we have
\begin{eqnarray*}
\left\|\zeta(t,\zeta_0)-p(t)\right\|\leq \left\| \zeta_0  \right\| + \displaystyle \int_{0}^{t} L_f \left\| \zeta(s,\zeta_0)-p(s) \right\| ds. 
\end{eqnarray*}

Implementing Gronwall-Bellman Lemma \cite{Corduneanu77} to the last inequality and using $\left\|\zeta_0\right\|<\delta,$ one can see  for $t \in [0,2T]$ that the inequality
\begin{eqnarray*} 
\left\|\zeta(t,\zeta_0)-p(t)\right\| < \delta e^{2L_fT}
\end{eqnarray*}
is valid. Therefore, $\left\|\zeta(t,\zeta_0)-p(t)\right\|< \epsilon_1$ on the time interval $[0,2T],$ and consequently $\zeta(t,\zeta_0)$ intersects the surface $S$ at some moment $t_1(\zeta_0) \in [0,2T].$ In other words, the point $\zeta_1=\zeta(t_1(\zeta_0),\zeta_0)$ belongs to the surface $S.$  It is clear that $\left\|\zeta_1\right\| < R,$ where $R= \epsilon_1 + \rho.$ 
 
Now, let us fix an arbitrary number $l \in (0,1).$ By Lemma \ref{lemma1}, there exists a natural number $n_0=n_0\left(\frac{l\delta}{R}\right),$ which is independent of $\zeta_0,$ such that 
\begin{eqnarray}\label{boundedness_2}
\begin{array}{l}
\left\|\zeta(n_0 T+t_1(\zeta_0),\zeta_0)\right\| 
= \left\|\zeta(n_0T,\zeta_1)\right\| < \displaystyle \frac{l\delta}{R} \left\|\zeta_1\right\| < l\delta.
\end{array}
\end{eqnarray}

Let us take $\epsilon = \displaystyle \left(\frac{1-l}{2}\right)\delta$ and choose a nonzero number $\mu$ such that $\left|\mu\right|<\displaystyle \frac{\epsilon L_f}{M_g \left[e^{L_f(n_0+2)T}-1\right]}.$

Fix an arbitrary solution $\eta_{x(t)}(t,\eta_0),$ where $x(t)$ is a chaotic solution of system $(\ref{1})$ and $\eta_0 \in B_{\delta}$ with the number $\delta$ as specified above. 
Herewith, there exists a number $t_1(\eta_0) \in [0,2T]$ such that $\zeta(t_1(\eta_0),\eta_0)$ belongs to the surface $S.$

Making use of the integral equations
\begin{displaymath}
\eta_{x(t)}(t,\eta_0)=\eta_0 + \displaystyle \int_{0}^{t} f(\eta_{x(t)}(s,\eta_0)) ds 
+ \mu \displaystyle \int_{0}^{t} g(x(s)) ds
\end{displaymath}
and
\begin{displaymath}
\zeta(t,\eta_0)=\eta_0 + \displaystyle \int_{0}^{t} f(\zeta(s,\eta_0)) ds 
\end{displaymath}
one can obtain for $t \in [0, (n_0+2)T]$ that
\begin{eqnarray*}
\left\|\eta_{x(t)}(t,\eta_0)-\zeta(t,\eta_0)\right\| \leq  \left|\mu\right| M_g t  + \displaystyle \int_{0}^{t} L_f \left\| \eta_{x(t)}(s,\eta_0)- \zeta(s,\eta_0) \right\| ds. 
\end{eqnarray*}

Application of Lemma $2.2$ \cite{Bar70} to the last inequality implies that
\begin{eqnarray*}
&& \left\|\eta_{x(t)}(t,\eta_0)-\zeta(t,\eta_0)\right\|  \leq  \left|\mu\right| M_g t + \left|\mu\right| M_g  L_f \displaystyle \int_{0}^{t} s e^{L_f(t-s)}   ds \\
&& = \frac{\left|\mu\right|M_g}{L_f} \left(e^{L_f t}-1\right)\\
&& \leq  \frac{\left|\mu\right|M_g}{L_f} \left(e^{L_f (n_0+2)T}-1\right)\\
&& < \epsilon,
\end{eqnarray*}
for all $t \in [0, (n_0+2)T].$

In this case, we achieve the inequality
\begin{eqnarray*}
\left\| \eta_{x(t)}(n_0T+t_1(\eta_0),\eta_0) - \zeta(n_0T+t_1(\eta_0),\eta_0) \right\| < \epsilon,
\end{eqnarray*}
and hence by means of $(\ref{boundedness_2})$ we get
\begin{eqnarray*}
&& \left\| \eta_{x(t)}(n_0T+t_1(\eta_0),\eta_0)  \right\|  \leq \left\| \eta_{x(t)}(n_0T+t_1(\eta_0),\eta_0) - \zeta(n_0T+t_1(\eta_0),\eta_0) \right\|  + \left\| \zeta(n_0T+t_1(\eta_0),\eta_0) \right\| \\ 
&& < \epsilon + l\delta\\
&& < \delta. 
\end{eqnarray*}
In other words, $\eta_1=\eta_{x(t)}(\theta_1,\eta_0) \in B_{\delta},$ where $\theta_1=n_0T+t_1(\eta_0).$ We note that the point $\eta_1$ depends on both the initial data $\eta_0 \in B_{\delta}$ and the chaotic function $x(t).$  

Performing a similar procedure as presented above, by means of the solution $\zeta(t-\theta_1,\eta_1)$ of system $(\ref{3}),$ one can obtain that the inequality
\begin{displaymath}
\left\|\eta_{x(t)}(t,\eta_0)-\zeta(t-\theta_1,\eta_1)\right\| \leq  \frac{\left|\mu\right|M_g}{L_f} \left(e^{L_f (n_0+2)T}-1\right)
\end{displaymath}
holds for all $t \in [\theta_1, \theta_1+(n_0+2)T].$ On the other hand, the existence of a number $t_2(\eta_1)\in [0,2T]$ such that $\zeta(t_2(\eta_1),\eta_1) \in S$ is provable.

Therefore we have
\begin{displaymath}
\left\| \eta_{x(t)}(2n_0T+t_1(\eta_0)+t_2(\eta_1),\eta_0) - \zeta(n_0T+t_2(\eta_1),\eta_1) \right\| < \epsilon,
\end{displaymath}
and hence $ \eta_2=\eta_{x(t)}(\theta_2,\eta_0) \in B_{\delta},$ where $\theta_2=2n_0T+t_1(\eta_0)+t_2(\eta_1).$

\begin{figure*}[t] 
\centering
\includegraphics[width=11.0cm]{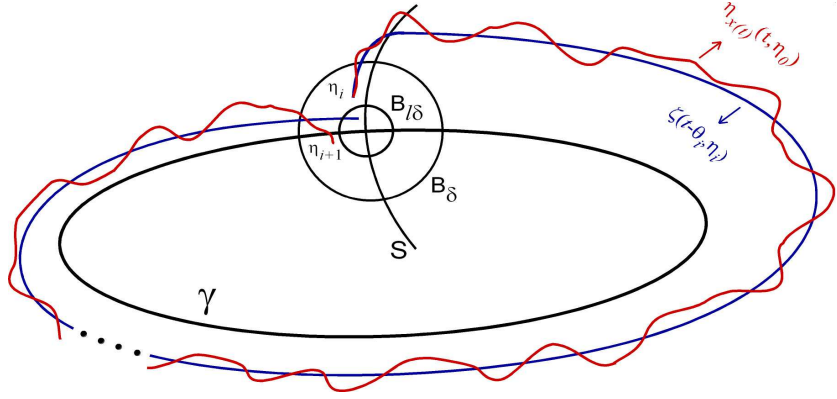}
\caption{The representational visualization of the proof of Theorem \ref{boundedness}. The trajectory in red color shows the solution $\eta_{x(t)}(t,\eta_0)$ of system $(\ref{2}),$ while the trajectory in blue color represents the solution $\zeta(t-\theta_i,\eta_i)$ of system $(\ref{3}),$ where the sequence $\left\{\theta_i\right\}$ is defined through formula $(\ref{formula_sequence})$ and $\eta_i=\eta_{x(t)}(\theta_i,\eta_0).$  The presented illustration covers the way of the $2-$dimensional case of the proof on the time interval $[\theta_i,\theta_{i+1}],$ for an arbitrary nonnegative integer $i.$ At the moment $t=\theta_{i+1},$ the solution $\zeta(t-\theta_i,\eta_i)$ belongs to the ball $B_{l\delta},$ and on the other hand, $\eta_{i+1}$ is inside the ball $B_{\delta}.$ }
\label{picture}
\end{figure*}

One can continue in the same manner to constitute a sequence $\left\{t_j\right\}, $ which depends both on the initial condition $\eta_0 \in B_{\delta}$ and the chaotic function $x(t),$ and satisfies $0\leq t_j \leq 2T,$ $j\geq 1,$ such that for each integer $i\geq 0$ we have  $\eta_i=\eta_{x(t)}(\theta_i,\eta_0) \in B_{\delta}$ and
\begin{eqnarray}
\begin{array}{l}\label{cyclic_motion}
\left\|\eta_{x(t)}(t,\eta_0)-\zeta(t-\theta_i,\eta_i)\right\| \leq  \displaystyle \frac{\left|\mu\right|M_g}{L_f} \left(e^{L_f (n_0+2)T}-1\right),
\end{array}
\end{eqnarray}
where the sequence $\left\{\theta_i\right\}$ is defined through the equation 
\begin{eqnarray}\label{formula_sequence}
\theta_i=in_0T+\sum_{j=1}^i t_j, ~i\geq 1,
\end{eqnarray} 
and $\theta_0=0.$  We emphasize that for any $i\geq 1,$ it is true that $\theta_i \in \left[ in_0T,i(n_0+2)T  \right]$ and  $\theta_i-\theta_{i-1}=n_0T+t_i\leq (n_0+2)T.$  The procedure of the proof for $t \in [\theta_i,\theta_{i+1}]$ is illustrated in Figure $\ref{picture}.$

In the remaining part of the proof, we will indicate boundedness of the function $\eta_{x(t)}(t,\eta_0)-p(t),$ which implies boundedness of $\eta_{x(t)}(t,\eta_0).$ We note that $\eta_{x(t)}(t,\eta_0)-p(t)$ is a solution of system $(\ref{6}),$ satisfying the initial condition $\eta_{x(t)}(0,\eta_0)-p(0)=\eta_0.$

Using the relations
\begin{displaymath}
\zeta(t-\theta_j,\eta_j)=\eta_j + \displaystyle \int_{\theta_j}^{t} f(\zeta(s-\theta_j,\eta_j))ds 
\end{displaymath}
and 
\begin{displaymath}
p(t)= p(\theta_j) + \displaystyle \int_{\theta_j}^{t} f(p(s))ds 
\end{displaymath}
on the time interval $[\theta_j,\theta_j+2T],$  together with Gronwall-Bellman Lemma, we achieve the inequality  
\begin{displaymath}
\left\| \zeta(t-\theta_j,\eta_j)-p(t)  \right\| \leq (\delta+\rho) e^{2L_f T}.
\end{displaymath}
Hereby, one can see that 
\begin{eqnarray*}
&& \left\|\zeta(t_{j+1}(\eta_j),\eta_j)\right\|  \leq \left\| \zeta(t_{j+1}(\eta_j),\eta_j)-p(t_{j+1}(\eta_j))  \right\| + \left\|p(t_{j+1}(\eta_j)) \right\| \\
&& \leq \rho + (\delta+\rho) e^{2L_f T}.
\end{eqnarray*} 

Since the point $\zeta(t_{j+1}(\eta_j),\eta_j)$ is on the surface $S,$ according to inequality $(\ref{14}),$ for all $t\ge 0,$ we have 
\begin{eqnarray*}
&& \left\| \zeta(t+t_{j+1}(\eta_j),\eta_j) - p(t)\right\|  = \left\| \zeta(t, \zeta(t_{j+1}(\eta_j), \eta_j)) - p(t) \right\| \\
&& \leq  4K_1\left\|P^{-1}(0)\right\|   \left\| \zeta(t_{j+1}(\eta_j), \eta_j) \right\|   e^{-\alpha t/2} \\
&& \leq  4K_1\left\|P^{-1}(0)\right\|   \left[\rho + \left(\delta + \rho \right) e^{2L_f T} \right] e^{-\alpha t/2}
\end{eqnarray*} 
and therefore one attains for all $t \geq 0$ that 
\begin{eqnarray*}
\left\| \zeta(t+t_{j+1}(\eta_j),\eta_j) - p(t)\right\|  \leq 4K_1\left\|P^{-1}(0)\right\|  \left[\rho + \left(\delta + \rho \right) e^{2L_f T} \right].  
\end{eqnarray*}

Thus, for $t \in [\theta_j, \theta_{j+1}],$ we have 
\begin{eqnarray}
\begin{array}{l} \label{boundedness_1}
\left\|\eta_{x(t)}(t,\eta_0)-p(t)\right\| \leq  \left\|\eta_{x(t)}(t,\eta_0) -\zeta(t-\theta_j,\eta_j)\right\|  +\left\|\zeta(t-\theta_j,\eta_j)-p(t)\right\|  \\
\leq  \displaystyle \frac{\left|\mu\right|M_g}{L_f} \left(e^{L_f (n_0+2)T}-1\right) + H(\delta,\rho),
\end{array}
\end{eqnarray}
where $ H(\delta, \rho)$ is the maximum of the numbers $(\delta+\rho) e^{2L_f T}$ and $4K_1\left\|P^{-1}(0)\right\|  \left[\rho + \left(\delta + \rho \right) e^{2L_f T}\right].$ It is worth saying that $H(\delta, \rho) \to 0$ as $\delta \to 0$ and $\rho \to 0,$ and $\left\|\eta_{x(t)}(t,\eta_0)-p(t)\right\| $ can be made arbitrarily small by suitable choices of $\mu, \delta$ and $\rho.$ 

Consequently, any solution $\eta_{x(t)}(t,\eta_0)$ of system $(\ref{2})$ satisfying the condition $\eta_0 \in B_{\delta}$ is bounded on $\mathbb R_+.$

Proof of the theorem is finalized. $\square$

Now, we fix values of $\mu$ and $\delta$ as specified in Theorem  $\ref{boundedness},$ and use these values in the next section, where we will discuss the extension problem of sensitivity.

\section{Entrainment of sensitivity} \label{sec_sensitivity}

In the present section, extension of sensitivity feature through system  $(\ref{2})$ will be handled. We shall begin with the meaning of the aforementioned property for systems $(\ref{1})$ and $(\ref{2}),$ and then state the main theorem of the present section.

We say that system $(\ref{1})$ is sensitive if there exist positive real numbers $\epsilon_0$ and $\Delta$ such that for an arbitrary positive real number $\delta_0$ and for each chaotic solution $x(t)$ of system $(\ref{1}),$ there exist a chaotic solution $\overline{x}(t)$ of the same system and an interval $J \subset \mathbb R_+,$ with length not less than $\Delta,$ such that $\left\|x(0)-\overline{x}(0)\right\|<\delta_0$ and $\left\|x(t)-\overline{x}(t)\right\| > \epsilon_0,$ for all $t \in J.$ 

Now, suppose that $\delta$ is a positive number as indicated in Theorem $\ref{boundedness}.$ In a similar way, the system $(\ref{2})$ is called sensitive if there exist positive real numbers $\epsilon_1$ and $\overline{\Delta}$ such that for an arbitrary positive number $\delta_1<\delta$ and for each solution $\eta_{x(t)}(t,\eta_0),$ where $x(t)$ is a chaotic function and $\eta_0 \in B_{\delta},$  there exist an interval $J^1 \subset J,$ with length not less than $\overline{\Delta},$ and a solution $\eta_{\overline{x}(t)}(t,\eta_1),$ where $\overline{x}(t)$ is a chaotic function and $\eta_1 \in B_{\delta},$ such that $\left\|\eta_0-\eta_1\right\|<\delta_1$ and $\left\|\eta_{x(t)}(t,\eta_0)-\eta_{\overline{x}(t)}(t,\eta_1)\right\| > \epsilon_1,$ for all $t \in J^1.$ 

Through the next theorem, the extension of sensitivity feature from system $(\ref{1})$ to system $(\ref{2})$ will be mentioned. 

In addition, we shall say that system $(\ref{1})+(\ref{2})$ is sensitive if  systems $(\ref{1})$ and $(\ref{2})$ are both sensitive. This description is a natural one since, otherwise, the inequality 
$\left\| x(t) - \overline{x}(t)\right\| > \epsilon_0$ implies that $\left\| \left(x(t), \eta_{x(t)}(t)\right) - \left(\overline{x}(t),\eta_{\overline{x}(t)}(t) \right) \right\| > \epsilon_0$ in the same interval of time, which already signifies unpredictability of system $(\ref{1})+(\ref{2}).$ In our theory of entrainment of chaos the crucial idea is  not only the extension of sensitivity through system $(\ref{1})+(\ref{2}),$ but also through system $(\ref{2})$ individually. For this reason, one should understand sensitivity of system $(\ref{1})+(\ref{2})$ as a property which is equivalent to the sensitivity of system $(\ref{2}).$

\begin{theorem} \label{sensitivity_thm}
If system $(\ref{1})$ is sensitive then the same is true for system $(\ref{2}).$
\end{theorem}

\noindent \textbf{Proof.}
Fix an arbitrary positive number $\delta_1<\delta$ and let $\eta_{x(t)}(t,\eta_0)$ be a given solution of system $(\ref{2}),$ where $x(t)$ is a chaotic function and $\eta_0 \in B_{\delta}.$ Since system $(\ref{1})$ is sensitive, there exist positive real numbers $\epsilon_0$ and $\Delta$ such that for any fixed number $\delta_0>0$ the inequalities $\left\|x(0)-\overline{x}(0)\right\|<\delta_0$ and $\left\|x(t)-\overline{x}(t)\right\| > \epsilon_0,$ $t \in J,$ hold for some chaotic solution $\overline{x}(t)$ of system $(\ref{1})$ and for some interval $J \subset \mathbb R_+,$ whose length is not less than $\Delta.$ 

Now, let us fix $\eta_1 \in B_{\delta}$ such that $\left\|\eta_0-\eta_1\right\|<\delta_1.$ In the proof our aim is to determine positive real numbers $\epsilon_1$ and $\overline{\Delta}$ such that for some interval $J^1\subset J$ with length $\overline{\Delta},$ the inequality $\left\|\eta_{x(t)}(t,\eta_0)-\eta_{\overline{x}(t)}(t,\eta_1)\right\| > \epsilon_1,$ holds for all $t \in J^1.$ 

Suppose that
$
 g(x)
 = \left( \begin{array}{ccc}
 g_1(x)   \\
 g_2(x) \\
\vdots \\
 g_n(x)
\end{array} \right),
$
where each $g_j,$ $1 \leq j \leq n,$ is a real valued function.

Since for each chaotic solution $x(t)$ of system $(\ref{1})$ the function $x'(t)$ is inside the tube with radius $M_F,$ one can conclude that the collection of chaotic solutions of system $(\ref{1})$ constitute an equicontinuous family on $\mathbb R_+.$ Making use of the uniform continuity of the function $\overline{g}: \mathbb R^m \times \mathbb R^m  \to \mathbb R^n,$ defined as $\overline{g}(x_1,x_2)=g(x_1)-g(x_2),$ on the compact region 
\begin{eqnarray*}
\mathscr{D}=\left\{(x_1,x_2) \in \mathbb R^m \times \mathbb R^m  ~|~ \left\|x_1\right\| \leq M, \left\|x_2\right\| \leq M \right\}, \nonumber 
\end{eqnarray*}
together with the equicontinuity of the collection of chaotic solutions of system $(\ref{1}),$  one can verify that the set consisting of the elements of the form
$g_j(x(t))-g_j(\overline{x}(t)), 1\leq j\leq n,$ where $x(t)$ and $\overline{x}(t)$ are chaotic solutions of system $(\ref{1}),$ is an equicontinuous family on $\mathbb R_+.$

Therefore, there exists a positive real number $\tau<\Delta,$ independent of the functions $x(t)$ and $\overline{x}(t),$  such that for any $t_1,t_2\in \mathbb R_+$ with $\left|t_1-t_2\right|<\tau,$ the inequality 
\begin{eqnarray} \label{sensitivity_proof_1}
\begin{array}{l}
\left| \left(g_j\left(x(t_1)\right) - g_j\left(\overline{x}(t_1)\right)  \right) - \left(g_j\left(x(t_2)\right) - g_j\left(\overline{x}(t_2)\right)  \right)   \right| <\displaystyle \frac{L_g\epsilon_0}{2n}
\end{array}
\end{eqnarray}
holds for all $1\leq j \leq n.$

Condition $(A2)$ implies for all $t \in J$ that $\left\|g(x(t))-g(\overline{x}(t)) \right\| \geq L_g \left\|x(t)-\overline{x}(t)\right\|.$   Therefore, for each $t\in J,$ there exists an integer $j_0,$ $1 \leq j_0 \leq n,$ which possibly depends on $t,$ such that 
\begin{eqnarray}
\begin{array}{l}
\left|g_{j_0}(x(t))-g_{j_0}(\overline{x}(t))\right|   \geq \displaystyle \frac{L_g}{n} \left\|x(t)-\overline{x}(t)\right\| \nonumber.
\end{array}
\end{eqnarray}
Otherwise, if there exists $s\in J$ such that for all $1\leq j\leq n,$ the inequality  
\begin{eqnarray}
\begin{array}{l}
\left|g_{j} \left(x\left(s\right)  \right)-g_{j}(\overline{x}(s)) \right|< \displaystyle \frac{L_g}{n} \left\|x(s)-\overline{x}(s)\right\| \nonumber
\end{array}
\end{eqnarray}
holds, then one encounters with a contradiction since
\begin{eqnarray*}
\left\|g(x(s))-g(\overline{x}(s))  \right\|   \leq \sum_{j=1}^{n}\left| g_{j}(x(s))-g_{j}(\overline{x}(s)) \right|  < L_g \left\|x(s)-\overline{x}(s)\right\|. \nonumber
\end{eqnarray*}

Now, let $s_0$ be the midpoint of the interval $J$ and $\theta=s_0-\frac{\tau}{2}.$ One can find an integer $j_0=j_0(s_0),$  $1 \leq j_0 \leq n,$ such that 
\begin{eqnarray}
\begin{array}{l} \label{sensitivity_proof_2}
\left|g_{j_0}(x(s_0))-g_{j_0}(\overline{x}(s_0))\right|   \geq \displaystyle\frac{L_g}{n} \left\|x(s_0)-\overline{x}(s_0)\right\| > \displaystyle\frac{L_g\epsilon_0}{n}. 
\end{array}
\end{eqnarray}

On the other hand, making use of the inequality $(\ref{sensitivity_proof_1}),$ for all $t \in \left[\theta, \theta+\tau\right]$ we have
\begin{eqnarray*}
&& \left|g_{j_0}\left(x(s_0)\right) - g_{j_0}\left(\overline{x}(s_0)\right) \right| - \left|g_{j_0}\left(x(t)\right) - g_{j_0}\left(\overline{x}(t)\right) \right| \\
&& \leq \left| \left(g_{j_0}\left(x(t)\right) - g_{j_0}\left(\overline{x}(t)\right)  \right) - \left(g_{j_0}\left(x(s_0)\right) - g_{j_0}\left(\overline{x}(s_0)\right)  \right)   \right| \\
&&<\frac{L_g\epsilon_0}{2n}
\end{eqnarray*}
and therefore, by means of $(\ref{sensitivity_proof_2}),$ we achieve that the inequality
\begin{eqnarray} \label{sensitivity_proof_3}
\begin{array}{l} 
 \left|g_{j_0}\left(x(t)\right) - g_{j_0}\left(\overline{x}(t)\right) \right| > \left|g_{j_0}\left(x(s_0)\right) - g_{j_0}\left(\overline{x}(s_0)\right) \right|  - \displaystyle \frac{L_g\epsilon_0}{2n} > \displaystyle \frac{L_g\epsilon_0}{2n}
 \end{array}
\end{eqnarray}
is valid for all $t\in \left[\theta, \theta+\tau\right].$

Making use of the mean value theorem for integrals, one can find numbers $s_1, s_2, \ldots, s_n \in [\theta,\theta+\tau]$ such that
\begin{eqnarray*}
&& \left\|\displaystyle\int^{\theta + \tau}_{\theta} \left[g(x(s))-g(\overline{x}(s))\right] ds \right\| \nonumber\\
&& = \left\|\left( \begin{array}{ccc}
\displaystyle\int^{\theta + \tau}_{\theta} \left[g_1(x(s))-g_1(\overline{x}(s))\right] ds  \\
\displaystyle\int^{\theta + \tau}_{\theta} \left[g_2(x(s))-g_2(\overline{x}(s))\right] ds \\
\vdots \\
\displaystyle\int^{\theta + \tau}_{\theta} \left[g_n(x(s))-g_n(\overline{x}(s))\right] ds \\
\end{array} \right)\right\| \\
&& = \left\|\left( \begin{array}{ccc}
\tau \left[g_1(x(s_1))-g_1(\overline{x}(s_1))\right]   \\
\tau \left[g_2(x(s_2))-g_2(\overline{x}(s_2))\right]  \\
\vdots \\
\tau \left[g_n(x(s_n))-g_n(\overline{x}(s_n))\right] 
\end{array} \right)\right\|.
\end{eqnarray*}

Thus, using the inequality $(\ref{sensitivity_proof_3}),$ we attain that
\begin{eqnarray} \label{sensitivity_proof_4}
\begin{array}{l} 
\left\|\displaystyle\int^{\theta + \tau}_{\theta} \left[g(x(s))-g(\overline{x}(s))\right] ds \right\| \geq \tau  \left|g_{j_0}(x(s_{j_0}))-g_{j_0}(\overline{x}(s_{j_0}))\right| > \displaystyle \frac{\tau  L_g \epsilon_0}{2n}.
\end{array}
\end{eqnarray}

For $t\in [\theta,\theta+\tau],$ the solutions $\eta_{x(t)}(t,\eta_0)$ and $\eta_{\overline{x}(t)}(t,\eta_1)$ satisfy the integral equations
\begin{eqnarray*}
&& \eta_{x(t)}(t,\eta_0)= \eta_{x(t)}(\theta,\eta_0)+ \displaystyle\int^{t}_{\theta} f(\eta_{x(t)}(s,\eta_0)) ds  + \displaystyle\int^{t}_{\theta} \mu g(x(s)) ds,  
\end{eqnarray*}
and
\begin{eqnarray*}
&& \eta_{\overline{x}(t)}(t,\eta_1)= \eta_{\overline{x}(t)}(\theta,\eta_1)+ \displaystyle\int^{t}_{\theta} f(\eta_{\overline{x}(t)}(s,\eta_1)) ds  + \displaystyle\int^{t}_{\theta} \mu g(\overline{x}(s)) ds, 
\end{eqnarray*}
respectively, and herewith the equation
\begin{eqnarray*}
&& \eta_{x(t)}(t,\eta_0)-\eta_{\overline{x}(t)}(t,\eta_1)  = (\eta_{x(t)}(\theta,\eta_0)-\eta_{\overline{x}(t)}(\theta,\eta_1))  \\
&& + \displaystyle\int^{t}_{\theta}  \left[ f(\eta_{x(t)}(s,\eta_0))-f(\eta_{\overline{x}(t)}(s,\eta_1)) \right]   ds  \\
&& + \displaystyle\int^{t}_{\theta} \mu [g(x(s))-g(\overline{x}(s))]  ds
\end{eqnarray*}
is achieved.
Hence, we have the inequality
\begin{eqnarray} \label{sensitivity_proof_5}
\begin{array}{l}
\left\|\eta_{x(t)}(\theta+\tau,\eta_0)-\eta_{\overline{x}(t)}(\theta+\tau,\eta_1) \right\| \geq  \left|\mu\right| \left\|\displaystyle\int^{\theta+\tau}_{\theta}  [g(x(s))-g(\overline{x}(s))] ds \right\| \\
- \left\| \eta_{x(t)}(\theta,\eta_0)-\eta_{\overline{x}(t)}(\theta,\eta_1) \right\| - \displaystyle\int^{\theta+\tau}_{\theta}  L_f\left\| \eta_{x(t)}(s,\eta_0)-\eta_{\overline{x}(t)}(s,\eta_1)\right\| ds.
\end{array}
\end{eqnarray}

Now, assume that 
\begin{eqnarray*}
\displaystyle \max_{t\in [\theta,\theta+\tau]}\left\| \eta_{x(t)}(t,\eta_0)-\eta_{\overline{x}(t)}(t,\eta_1)\right\| \leq \frac{\left|\mu\right|\tau L_g \epsilon_0}{2n(2+\tau L_f)}.
\end{eqnarray*}
In this case, one arrives at a contradiction since, by means of the inequalities $(\ref{sensitivity_proof_4})$ and $(\ref{sensitivity_proof_5}),$ we have 
\begin{eqnarray*}
&& \displaystyle \max_{t\in [\theta,\theta+\tau]}\left\| \eta_{x(t)}(t,\eta_0)-\eta_{\overline{x}(t)}(t,\eta_1)\right\|  \geq \left\|\eta_{x(t)}(\theta+\tau,\eta_0)-\eta_{\overline{x}(t)}(\theta+\tau,\eta_1)\right\| \\
&& > \frac{\left|\mu\right|\tau L_g \epsilon_0}{2n}  - (1+ \tau L_f) \displaystyle \max_{t\in [\theta,\theta+\tau]}\left\|\eta_{x(t)}(t,\eta_0)-\eta_{\overline{x}(t)}(t,\eta_1)\right\| \\
&& \geq \frac{\left|\mu\right|\tau L_g \epsilon_0}{2n} - (1+ \tau L_f) \frac{\left|\mu\right|\tau L_g \epsilon_0}{2n(2+\tau L_f)} \\
&& = \frac{\left|\mu\right|\tau L_g \epsilon_0}{2n} \left(1-\frac{1+ \tau L_f}{2+ \tau L_f}\right) \\
&& = \frac{\left|\mu\right|\tau L_g \epsilon_0}{2n( 2+\tau L_f )}.
\end{eqnarray*}
Therefore, we have 
\begin{eqnarray*}
\displaystyle \max_{t\in [\theta,\theta+\tau]}\left\|\eta_{x(t)}(t,\eta_0)-\eta_{\overline{x}(t)}(t,\eta_1) \right\| > \frac{\left|\mu\right|\tau L_g \epsilon_0}{2n(2+\tau L_f)}.
\end{eqnarray*}

Suppose that the function $\left\|\eta_{x(t)}(t,\eta_0)-\eta_{\overline{x}(t)}(t,\eta_1)\right\|$ takes its maximum on the interval $[\theta,\theta+\tau]$ at the point $\xi,$ that is, 
\begin{eqnarray*}
&& \displaystyle \max_{t \in [\theta,\theta+\tau]} \left\|\eta_{x(t)}(t,\eta_0)-\eta_{\overline{x}(t)}(t,\eta_1)\right\|  = \left\|\eta_{x(t)}(\xi,\eta_0)-\eta_{\overline{x}(t)}(\xi,\eta_1)\right\|, 
\end{eqnarray*} 
for some $\theta \leq \xi \leq \theta+\tau.$

For $t\in [\theta,\theta+\tau],$ by favour of the integral equations
\begin{eqnarray*}
\eta_{x(t)}(t,\eta_0)= \eta_{x(t)}(\xi,\eta_0)+ \displaystyle\int^{t}_{\xi} f(\eta_{x(t)}(s,\eta_0)) ds  + \displaystyle\int^{t}_{\xi} \mu g(x(s)) ds,  
\end{eqnarray*}
and
\begin{eqnarray*}
\eta_{\overline{x}(t)}(t,\eta_1)= \eta_{\overline{x}(t)}(\xi,\eta_1)+ \displaystyle\int^{t}_{\xi} f(\eta_{\overline{x}(t)}(s,\eta_1)) ds  + \displaystyle\int^{t}_{\xi} \mu g(\overline{x}(s)) ds,  
\end{eqnarray*}
we obtain
\begin{eqnarray*}
&& \eta_{x(t)}(t,\eta_0)-\eta_{\overline{x}(t)}(t,\eta_1) = (\eta_{x(t)}(\xi,\eta_0)-\eta_{\overline{x}(t)}(\xi,\eta_1))  \\
&& + \displaystyle\int^{t}_{\xi} \left[f(\eta_{x(t)}(s,\eta_0))-f(\eta_{\overline{x}(t)}(s,\eta_1))\right] ds  + \displaystyle\int^{t}_{\xi} \mu [g(x(s))-g(\overline{x}(s))]  ds.
\end{eqnarray*}
Define
\begin{displaymath}
\tau^1=\min \displaystyle \left\{ \frac{\tau}{2}, \frac{\left|\mu\right|\tau L_g \epsilon_0}{8n(K_0L_f+M_g\left|\mu\right|)(2+\tau L_f)}   \right\}
\end{displaymath}
and let
\begin{displaymath}
\theta^1=\left\{\begin{array}{ll} \xi, & ~\textrm{if}~  \xi \leq \theta + \frac{\tau}{2}   \\
\xi - \tau^1, & ~\textrm{if}~  \xi > \theta + \frac{\tau}{2}  \\
\end{array} \right. .\nonumber
\end{displaymath}
We note that the interval $J^1=[\theta^1, \theta^1+\tau^1]$ is a subset of $[\theta,\theta+\tau]$ and hence a subset of $J.$

For $t\in J^1,$ we have
\begin{eqnarray*} 
&& \left\|\eta_{x(t)}(t,\eta_0)-\eta_{\overline{x}(t)}(t,\eta_1)\right\|  \geq  \left\|\eta_{x(t)}(\xi,\eta_0)-\eta_{\overline{x}(t)}(\xi,\eta_1)\right\| \\
&& - \left|  \displaystyle\int^{t}_{\xi} L_f \left\|\eta_{x(t)}(s,\eta_0)-\eta_{\overline{x}(t)}(s,\eta_1)\right\| ds   \right|    - \left|\mu\right|   \left|  \displaystyle\int^{t}_{\xi} \left\|  g(x(s))-g(\overline{x}(s))  \right\| ds  \right| \\
&& > \displaystyle \frac{\left|\mu\right|\tau L_g \epsilon_0}{2n(2+\tau L_f)} -2K_0 L_f \tau^1-2M_g\left|\mu\right|\tau^1 \\
&& = \displaystyle\frac{\left|\mu\right|\tau L_g \epsilon_0}{2n(2+\tau L_f)} -2\tau^1 \left(K_0L_f+M_g\left|\mu\right|\right) \\
&& \geq \displaystyle\frac{\left|\mu\right|\tau L_g \epsilon_0}{4n(2+\tau L_f)}.
\end{eqnarray*}

Consequently, we achieve for $t \in J^1$ that
\begin{eqnarray} \label{sensitivity_ineq}
\left\|\eta_{x(t)}(t,\eta_0)-\eta_{\overline{x}(t)}(t,\eta_1)\right\| > \epsilon_1,
\end{eqnarray} 
where $\epsilon_1=\displaystyle \frac{\left|\mu\right|\tau L_g \epsilon_0}{4n(2+\tau L_f)}$ and the length $\tau^1$ of the interval $J^1$ does not depend on the chaotic functions $x(t)$ and $\overline{x}(t).$

The theorem is proved.$\square$

Thus the important property of sensitivity is proved. This property can be considered as the unique ingredient of chaos for a set of bounded solutions \cite{Rob95,Wiggins88}.

The formula $(\ref{cyclic_motion})$ provides us a support that chaotic solutions are near solutions of non-perturbed cyclically behaved motions and consequently, for sufficiently small $\left|\mu\right|,$ the system admits chaotic cycles. In the same time, in general, we do not request necessarily appearance of cyclical properties for chaotic solutions. That is, if $\left|\mu\right|$ is not sufficiently small, then chaotic solutions may not be cyclical. So, our results provide this possibility also.

Now, our aim is to exemplify our theoretical results through the illustration of $2-$dimensional projections of the chaotic attractor and the Poincar$\acute{e}$ section of a sample $4-$dimensional system which is in the form of system $(\ref{1})+(\ref{2}).$ We will  also verify through simulations the existence of the sensitivity feature in accordance with Theorem $\ref{sensitivity_thm}$. 

%Now, let us give an example of a system which satisfies the conditions of Andronov-Witt Theorem. 
For our purposes, we shall take into account the following $2-$dimensional system
\begin{eqnarray} \label{orbitally_stable_system}
\begin{array}{l}
u'_1=\alpha u_1-u_2-u_1(u_1^2+u_2^2) \\
u'_2=u_1+\alpha u_2- u_2(u_1^2+u_2^2),
\end{array}
\end{eqnarray}
which is in the form of system $(\ref{3}),$ where $\alpha$ is a positive real number and 
\begin{eqnarray} 
f(u_1,u_2) = \left( \begin{array}{ccc}
\alpha u_1-u_2-u_1(u_1^2+u_2^2)   \\
u_1+\alpha u_2- u_2(u_1^2+u_2^2)
\end{array} \right).
\end{eqnarray}

One can verify in this case that
$ 
p(t) = \left( \begin{array}{ccc}
\sqrt{\alpha}  \displaystyle \cos t  \\
\sqrt{\alpha} \displaystyle \sin t
\end{array} \right)
$
is a periodic solution of system $(\ref{orbitally_stable_system}).$ Evaluating  $A(t)=f'(p(t)),$ we achieve that
\begin{eqnarray} 
A(t)= \left(
\begin {array}{ccc}  
-2\alpha \displaystyle \cos^2 t & -1-\alpha \displaystyle \sin (2t) \label{matrix_A}\\
\noalign{\medskip}
1-\alpha \displaystyle \sin (2t) & -2\alpha\displaystyle \sin^2 t
\end {array}
\right).
\end{eqnarray}
Since the variational system $(\ref{4}),$ where the matrix function $A(t)$ is given by formula $(\ref{matrix_A}),$ admits the number $\rho_1=1$ as a multiplier, Lemma $7.3$ \cite{Hale80} implies that the second multiplier is given by $\rho_2=\exp \left(\int_0^{2\pi} tr A(s)ds\right) = e^{-4\pi \alpha}.$ Thus,  the second multiplier of the variational system corresponding to system $(\ref{orbitally_stable_system})$ is  in modulus less than one, and consequently the periodic solution $p(t)$ of this system is asymptotically orbitally  stable according to Andronov-Witt Theorem.

We take into consideration the chaotic Duffing's oscillator \cite{Th02}, as the generator of chaos, presented by the differential equation
\begin{eqnarray} 
\begin{array}{l}
x''+0.05x'+x^3=7.5\displaystyle \cos t.\label{ueda_1}
\end{array}
\end{eqnarray}
Defining the variables $x_1=x$ and $x_2=x',$ equation $(\ref{ueda_1})$ can be rewritten as a system in the following form
\begin{eqnarray} 
\begin{array}{l}
x'_1=x_2 \\  \label{ueda_2}
x'_2=-0.05x_2-x_1^3+7.5\displaystyle \cos t.
\end{array}
\end{eqnarray}

In the remaining part of the example, we make use of the value $\alpha=9$ in system $(\ref{orbitally_stable_system}),$ and set up a unidirectional coupling between the systems $(\ref{ueda_2})$ and $(\ref{orbitally_stable_system})$ to achieve the following $4-$dimensional system
\begin{eqnarray} 
\begin{array}{l} \label{ueda_4}
x_1'=x_2,\\
x_2'=-0.05x_2-x_1^3+7.5\displaystyle \cos t,\\
x_3'=9x_3-x_4-x_3(x_3^2+x_4^2)+0.5x_1,\\
x_4'=x_3+9x_4-x_4(x_3^2+x_4^2)+3.6x_2.
\end{array}
\end{eqnarray}

\begin{figure}[ht] 
\centering
\includegraphics[width=10.5cm]{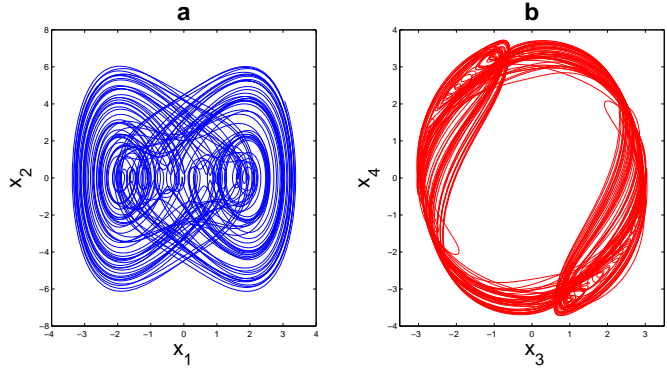}
\caption{The $2-$dimensional projections  of the chaotic attractor of system $(\ref{ueda_4}).$ (a) Projection on the $x_1-x_2$ plane, (b) Projection on the $x_3-x_4$ plane. The theoretical results indicate that system $(\ref{ueda_4})$ admits a chaotic attractor in the $4-$dimensional phase space. The picture in (a) represents not only the projection of the attractor of system $(\ref{ueda_4})$ on the $x_1-x_2$ plane, but also the chaotic attractor of system  $(\ref{ueda_2}).$ The chaotic attractor pictured in $(b),$ which appears around the limit cycle of system $(\ref{orbitally_stable_system})$ with $\alpha=9,$ is a manifestation of the entrainment mechanism. }
\label{entrainment1}
\end{figure}

We suppose that system $(\ref{ueda_4})$ admits a chaotic attractor in the $4-$dimensional phase space. In Figure $\ref{entrainment1},$ we visualize the $2-$dimensional projections on the $x_1-x_2$ and $x_3-x_4$ planes of the trajectory of system $(\ref{ueda_4})$ with initial data $x_1(0)=3.05, x_2(0)=4.153, x_3(0)=2.8, x_4(0)=0.5.$ The picture presented in Figure $\ref{entrainment1}, (a)$ is in fact the chaotic attractor of the system $(\ref{ueda_2})$ and  Figure $\ref{entrainment1}, (b)$ represents a chaotic attractor in a neighborhood of the limit cycle of system $(\ref{orbitally_stable_system}),$ with $\alpha=9.$ It is observable that the attractor around the limit cycle shown in Figure $\ref{entrainment1}, (b)$ repeated the chaotic structure of the Ueda's attractor presented in Figure $\ref{entrainment1}, (a).$

\begin{figure}[ht] 
\centering
\includegraphics[width=10.5cm]{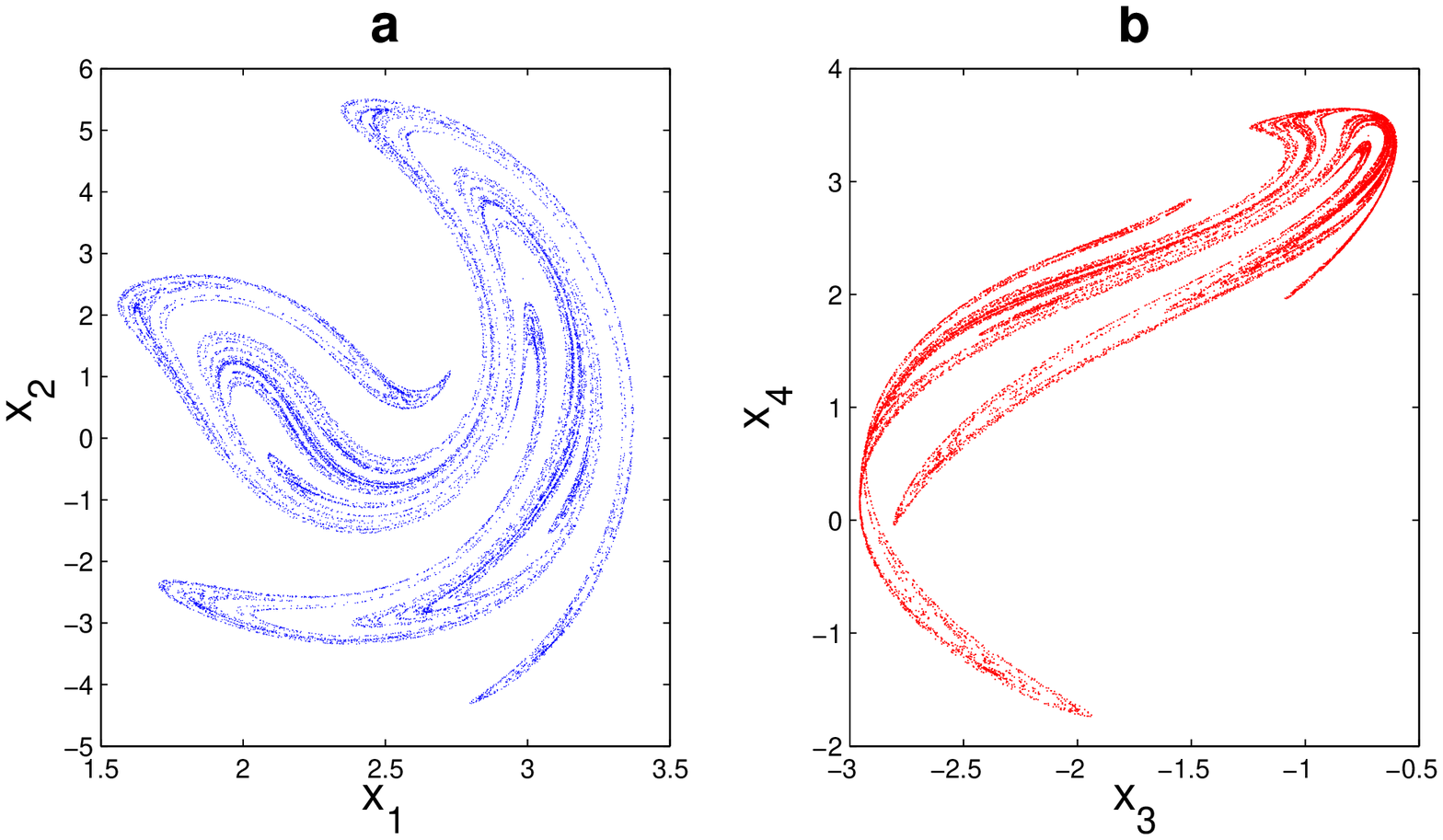}
\caption{The $2-$dimensional projections of the whole Poincar$\acute{e}$ section inside the $4-$dimensional phase space of system $(\ref{ueda_4}).$ (a) Projection on the $x_1-x_2$ plane, (b) Projection on the $x_3-x_4$ plane. The image shown in (a) is, in fact, the Ueda's strange attractor obtained through the Poincar$\acute{e}$ section of system $(\ref{ueda_2}).$  The picture in (b) represents a strange attractor around the limit cycle of system $(\ref{orbitally_stable_system}).$ Possibly one can call the strange attractor presented in (b) as the \textit{Ueda's strange attractor near limit cycle}.}
\label{entrainment2}
\end{figure}

Figure $\ref{entrainment2}$ pictures the $2-$dimensional projections of the whole Poincar$\acute{e}$ section inside the $4-$dimensional phase space of system $(\ref{ueda_4}),$ which is obtained by marking the trajectory  corresponding to initial data $x_1(0)=2, x_2(0)=3, x_3(0)=3, x_4(0)=0$ stroboscopically at times that are integer multiples of $2\pi.$  Figure $\ref{entrainment2},(a)$ represents the projection of the Poincar$\acute{e}$ section on the $x_1-x_2$ plane, and we note that this projection is, in fact, the strange attractor of system $(\ref{ueda_2}).$ On the other hand, the projection on the  $x_3-x_4,$ plane presented in Figure $\ref{entrainment2}, (b)$ is a strange attractor around the limit cycle of system $(\ref{orbitally_stable_system}).$ It is apparent that the attractor indicated in Figure $\ref{entrainment2}, (b)$ repeated the structure of the attractor shown in Figure $\ref{entrainment2},(a).$ The simulation results presented in Figure $\ref{entrainment1}$ and Figure $\ref{entrainment2}$ indicate the concept of entrainment of chaos.

\begin{figure}[ht] 
\centering
\includegraphics[width=10.5cm]{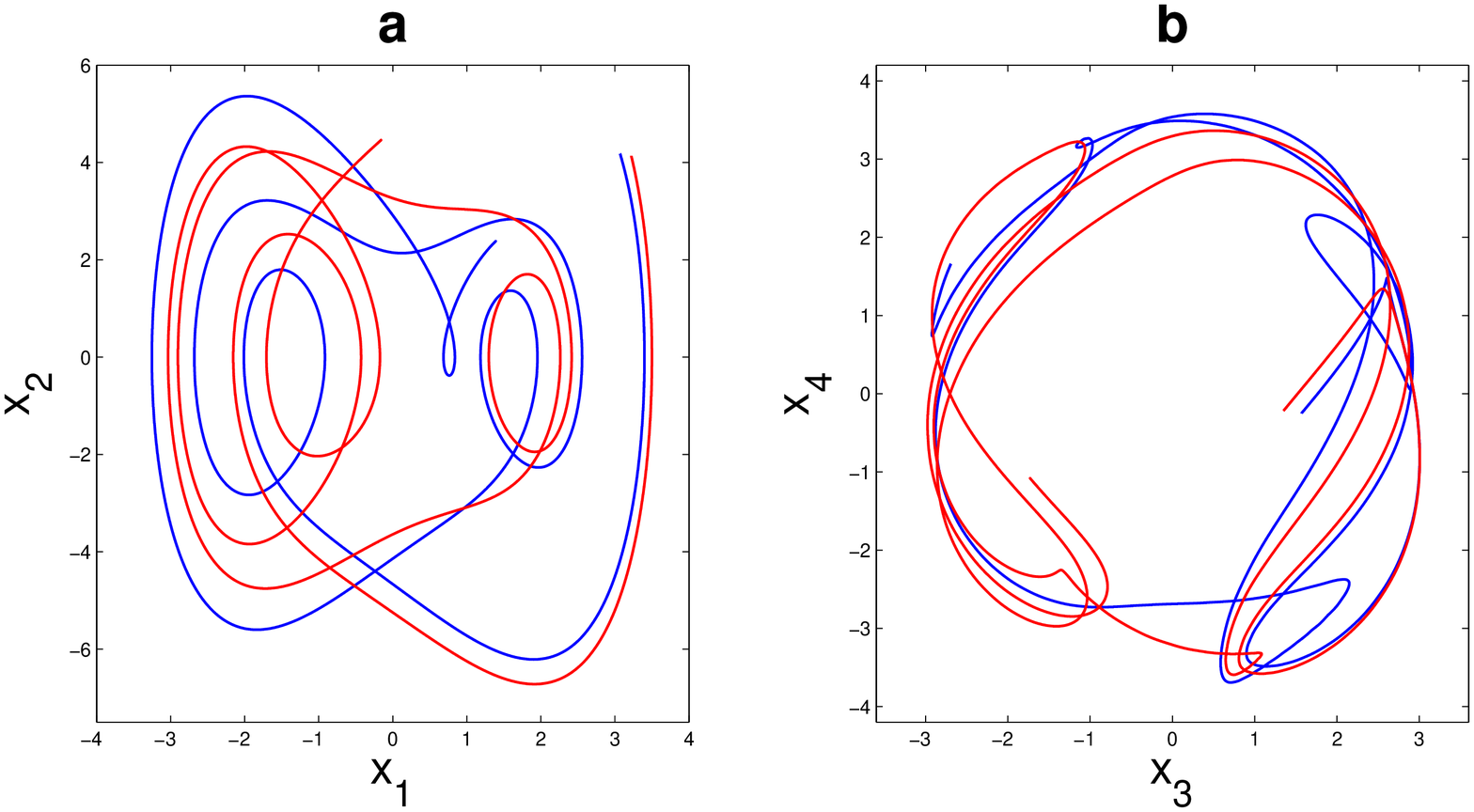}
\caption{Extension of sensitivity in system $(\ref{ueda_4}).$  (a)  Projection on the $x_1-x_2$ plane, (b) Projection on the $x_3-x_4$ plane. The picture in $(a),$ in fact, represents two initially nearby trajectories corresponding to system $(\ref{ueda_2}),$ which eventually diverge. The extension of sensitivity is observable in $(b)$ such that the blue and red trajectories are initially close to each other and then separated. It is noteworthy that since the chaotic perturbation in system $(\ref{ueda_4})$ is not weak enough  as indicated in Theorem $\ref{boundedness},$ it is visible in picture $(b)$ that the cyclic behavior is not achieved. The reason for the usage of a strong perturbation is to increase the strength of sensitivity, since otherwise, it is not able to visualize extension of this feature.}
\label{sensitivity_figure}
\end{figure}

Now, to show through simulations the extension of sensitivity, we consider two initially close solutions of system $(\ref{ueda_4}),$ one with the initial data $x_1(0)=3.07,  x_2(0)=4.18,  x_3(0)=1.57,  x_4(0)=-0.25,$ which is presented in blue color, and another with the initial data $x_1(0)=3.22,  x_2(0)=4.14,  x_3(0)=1.35,  x_4(0)=-0.22,$ which is pictured in red. In Figure $\ref{sensitivity_figure},$ we present $2-$dimensional projections of these trajectories on the $x_1-x_2$ and $x_3-x_4$ planes. The picture in Figure $\ref{sensitivity_figure}, (a)$ shows the sensitivity feature of system  $(\ref{ueda_2}),$ while picture in Figure $\ref{sensitivity_figure}, (b)$ represents the extension of this feature. 

We note that formula $(\ref{sensitivity_ineq})$ implies that the strength of sensitivity of system $(\ref{2})$ is proportional to the strength of the chaotic perturbation, $\mu g(x),$ used in this system. Therefore, if one considers system $(\ref{2})$ with weak perturbations, in that case, although the extension of sensitivity feature is guaranteed by Theorem $\ref{sensitivity_thm},$ it may not be visible in simulation results. But, according to formula $(\ref{cyclic_motion}),$ strong perturbations may diminish the cyclic behavior of the chaotic solutions. In this sense, the picture presented in Figure $\ref{sensitivity_figure}, (b)$ exhibits the extension of sensitivity, but does not indicate a cyclic behavior for the illustrated solution, according to the strength of the perturbation used in system  $(\ref{ueda_4}).$  

\section{Entrainment of period-doubling cascade} \label{sec_entrainment_period_doubling}

Period-doubling route to chaos is nowadays a popular way of indication of chaos \cite{Sander11}-\cite{Kovacic11} which has applications in  mechanics, electronics, biology, and is most convenient for simulations \cite{Parlitz85}-\cite{Zhang06}. 

We start this section by describing the chaos through period-doubling cascade for system $(\ref{1})$ and continue with its extension through system $(\ref{2}).$

To discuss the existence of chaos through period-doubling cascade, we consider the system
\begin{eqnarray}
\begin{array}{l}
x'=G(t,x,\lambda),  \label{period_doubling1}
\end{array}
\end{eqnarray}
where $\lambda$ is a parameter and the function $G: \mathbb R_+ \times \mathbb R^m \times \mathbb R \to \mathbb R^m$ satisfies the property that $F(t,x)=G(t,x,\lambda_{\infty}),$ for all $t\in\mathbb R_+, x\in\mathbb R^m$ and for some finite value $\lambda_{\infty}$ of the parameter $\lambda,$ which will be explained below. 

The following condition is needed throughout the section.

\begin{enumerate}
\item[\bf (A4)] There exists a positive real number $\omega$ such that the periodicity property $G(t+\omega,x,\lambda)=G(t,x,\lambda)$ holds, for all $t\in\mathbb R_+, x \in \mathbb R^m$ and $\lambda \in \mathbb R.$
\end{enumerate}

System $(\ref{1})$ is said to be chaotic through period-doubling cascade if there exist a natural number $k_0$ and a sequence of period-doubling bifurcation values $\left\{\lambda_j\right\},$  $\lambda_j \to \lambda_{\infty}$ as $j \to \infty,$ such that for each natural number $j$ as the parameter $\lambda$ increases or decreases through $\lambda_j,$ system $(\ref{period_doubling1})$ undergoes a period-doubling bifurcation and a periodic solution with period $k_02^j\omega$ appears. As a consequence, at the parameter value $\lambda=\lambda_{\infty},$ there exist infinitely many unstable periodic solutions of system $(\ref{period_doubling1}),$ and hence of system $(\ref{1}),$ all lying inside the tube with radius $M$. In this case, system $(\ref{1})$ admits periodic solutions of periods $k_0\omega, 2k_0\omega, 4k_0\omega, 8k_0\omega, \cdots.$ For details one can see \cite{Sander11,Feigenbaum80,Sander12}.  

In a similar way, we say that system $(\ref{2})$ admits the chaos through period-doubling cascade if system $(\ref{1})$ is chaotic through the same route and for each $k_0 2^j \omega-$periodic solution $x_j(t)$ of system $(\ref{1}),$ there exists a periodic solution $\eta_{x_j(t)}(t,\eta_0)$ of the system
\begin{eqnarray} \label{period_doubling2}
y'=f(y)+\mu g(x_j(t))
\end{eqnarray}
with the same period such that all such periodic solutions lie in a bounded region. Accordingly, system $(\ref{2})$ possesses infinitely many periodic solutions with periods $k_02^j\omega$ for each natural number $j$ such that all of these periodic solutions are unstable according to Theorem  $\ref{sensitivity_thm}.$

In the case that system $(\ref{2})$ is chaotic through period-doubling cascade as described above, we say that system $(\ref{1})+(\ref{2})$ admits the chaos through the same route and extension of this type of chaos is provided.

If system $(\ref{2})$ is chaotic through period-doubling cascade, in the same way as system $(\ref{period_doubling1})$, then system $(\ref{2})$ also undergoes period-doubling bifurcations as the parameter $\lambda$ increases or decreases through the values $\lambda_j.$ That is, the sequence $\left\{\lambda_j\right\}$ of bifurcation parameters is exactly the same for both systems. In the case system $(\ref{period_doubling1}),$ and consequently system $(\ref{1}),$ obey the Feigenbaum's universal behavior \cite{Th02,Zelinka,Feigenbaum80,Elaydi08,Sch05}, one can conclude that the same result holds, also, for system $(\ref{2})$. In other words, when $\lim_{j \to \infty } \frac{\lambda_j-\lambda_{j+1}}{\lambda_{j+1}-\lambda_{j+2}}$ is evaluated, the universal constant known as the Feigenbaum number $4.6692016\ldots$ is achieved  and this universal number is the same for systems $(\ref{1})$ and $(\ref{2})$, and consequently for system $(\ref{1})+(\ref{2})$.

Now, suppose that $x(t)$ is a $p_0-$periodic solution of system $(\ref{1})$ for some positive real number $p_0.$ Theorem $(\ref{boundedness})$ indicates that a bounded solution of the system $y'=f(y)+g(x(t))$ exists, provided that $\left|\mu\right|$ is sufficiently small. In the case that the dimension of system $(\ref{2})$ is $n=2,$ Massera's Theorem \cite{Massera50,Yoshizawa75} implies the existence of a $p_0-$periodic solution of the same system. On the other hand, by condition $(A2)$ the converse is also true. That is, if there exists a $p_0-$periodic solution of the system $y'=f(y)+g(x(t)),$ then $x(t)$ is necessarily $p_0-$periodic. Therefore, if system $(\ref{1})$ admits infinitely many periodic solutions with periods $k_02^j\omega$ for each natural number $j,$ then the same is true for system $(\ref{2})$ provided that $n=2.$ Additionally, no periodic solutions with any other period exist for this system. 

Next, we continue with an example which provides extension of chaos through period-doubling cascade. In paper \cite{Sato83}, it is mentioned that the Duffing's equation
\begin{eqnarray} 
\begin{array}{l} \label{per_doub_ex_1}
x''+0.3x'+x^3= \lambda \cos t, 
\end{array}
\end{eqnarray}
where $\lambda$ is a parameter, displays period-doubling bifurcations and admits the chaos through period-doubling cascade at the parameter value $\lambda=\lambda_{\infty}\equiv 40.$ Making use of the new variables $x_1=x$ and $x_2=x',$ we can rewrite equation $(\ref{per_doub_ex_1})$ as a system in the form
\begin{eqnarray}
\begin{array}{l} \label{perioddoubling_original}
x_1'=x_2, \\
x_2'=-0.3x_2-x_1^3+\lambda \displaystyle \cos t. 
\end{array}
\end{eqnarray}

To illustrate entrainment of chaos through period-doubling cascade, we shall make use of the parameter value $\lambda=\lambda_{\infty}$ in $(\ref{perioddoubling_original})$ and combine with system $(\ref{orbitally_stable_system})$ to constitute the $4-$dimensional system
\begin{eqnarray}
\begin{array}{l} \label{perioddoubling_example1}
x_1'=x_2, \\
x_2'=-0.3x_2-x_1^3+40 \displaystyle \cos t, \\
x_3'=\alpha x_3 -x_4 - x_3 \left(x_3^2+x_4^2\right) + \mu x_1, \\
x_4'= x_3 +\alpha x_4 - x_4 \left(x_3^2+x_4^2\right) + \mu x_2,
\end{array}
\end{eqnarray}
where $\mu$ is a non-zero real number and  $\alpha$ is a positive number.

We shall make use of the technique of Lyapunov functions to identify an annular region $G \subset \mathbb R^2$ which contains infinitely many periodic solutions of system 
\begin{eqnarray}
\begin{array}{l} \label{perioddoubling_example1a}
y_1'=\alpha y_1 -y_2 - y_1 \left(y_1^2+y_2^2\right) + \mu x_1, \\
y_2'= y_1 +\alpha y_2 - y_2 \left(y_1^2+y_2^2\right) + \mu x_2,
\end{array}
\end{eqnarray}
which is in the form of system $(\ref{2}),$ and this discussion will imply the existence of chaos through period-doubling cascade in that region. In such a case, the $2-$dimensional projection on the $x_3-x_4$ plane of the chaotic attractor of system $(\ref{perioddoubling_example1})$ will appear around the orbit $\gamma$ of the periodic solution $ 
p(t) = \left( \begin{array}{ccc}
\sqrt{\alpha}  \displaystyle \cos t  \\
\sqrt{\alpha} \displaystyle \sin t
\end{array} \right)
$ of system $(\ref{orbitally_stable_system}).$

Consider the Lyapunov function $V(x_3,x_4)=x_3^2+x_4^2.$
In this case one has
\begin{eqnarray*} 
&& V'_{(\ref{perioddoubling_example1})} (x_3,x_4) = 2 x_3 x'_3 + 2 x_4 x'_4 \\
&& = 2x_3 \left[ \alpha x_3 - x_4 -x_3   \left( x_3^2+x_4^2  \right) +\mu x_1   \right]   + 2x_4 \left[  x_3 + \alpha x_4 -x_4 \left( x_3^2+x_4^2  \right) +\mu x_2   \right] \\
&& = -2 \sqrt{x_3^2+x_4^2} \left[\left( x_3^2+x_4^2 - \alpha  \right)   \sqrt{x_3^2+x_4^2} - \mu \left(  \displaystyle \frac{x_1x_3}{\sqrt{x_3^2+x_4^2}}  +\displaystyle \frac{x_2x_4}{\sqrt{x_3^2+x_4^2}}  \right) \right].
\end{eqnarray*}

The inequality $\sqrt{x_3^2+x_4^2}> \sqrt{\alpha}+r_1,$ where $r_1$ is a positive number, implies that 
\begin{displaymath}
\sqrt{x_3^2+x_4^2} \left(x_3^2+x_4^2-\alpha\right)>r_1^3+3r_1^2\sqrt{\alpha}+2r_1 \alpha.
\end{displaymath}  
Similarly, if $0<r_2<\sqrt{x_3^2+x_4^2}<\sqrt{\alpha}-r_3,$ where $r_3<\sqrt{\alpha},$ then we have 
\begin{displaymath}
\sqrt{x_3^2+x_4^2} \left(x_3^2+x_4^2-\alpha\right) < r_2 r_3^2 -2r_2r_3\sqrt{\alpha}.
\end{displaymath}

Since the chaotic attractor of system $(\ref{perioddoubling_original})$ satisfies $\left|x_1\right|<6$ and $\left|x_2\right|<15,$ we attain that
\begin{eqnarray*}
\left| \mu \left(  \frac{x_1x_3}{\sqrt{x_3^2+x_4^2}}  +\frac{x_2x_4}{\sqrt{x_3^2+x_4^2}}  \right) \right| \leq 21 \left|\mu\right|.
\end{eqnarray*}

Therefore, if $\left|\mu\right|$ is sufficiently small such that 
\begin{eqnarray} \label{ineq_mu}
\left|\mu\right|<\frac{1}{21} \max \left\{ r_1^3, r_2r_3^2  \right\},
\end{eqnarray} 
then one can verify that 
$ V'_{(\ref{perioddoubling_example1})} (x_3,x_4)<0$ for $\sqrt{x_3^2+x_4^2}> \sqrt{\alpha}+r_1$ and $ V'_{(\ref{perioddoubling_example1})} (x_3,x_4)>0$ for $0<r_2<\sqrt{x_3^2+x_4^2}<\sqrt{\alpha}-r_3.$ Consequently, there are infinitely many periodic solutions inside the region
\begin{displaymath}
G=\left\{(x_3,x_4) \in \mathbb R^2:  \sqrt{\alpha}-r_3 \leq \sqrt{x_3^2+x_4^2} \leq \sqrt{\alpha} + r_1   \right\}.
\end{displaymath}  
It is worth saying that the region $G$ can be made arbitrarily narrow by small choices of the numbers $r_1,$ $r_3$ and $\left|\mu\right|$ according to inequality $(\ref{ineq_mu}).$
 
\begin{figure}[ht] 
\centering
\includegraphics[width=10.5cm]{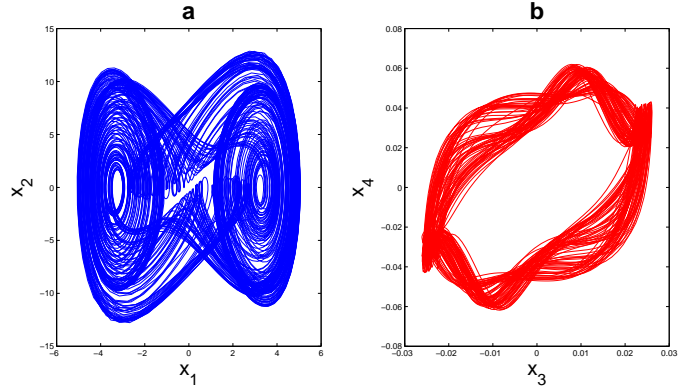}
\caption{$2-$dimensional projections of the chaotic attractor of system $(\ref{perioddoubling_example1}),$ where $\alpha=0.002$ and $\mu=0.008.$  (a)  Projection on the $x_1-x_2$ plane, (b) Projection on the $x_3-x_4$ plane.  The picture in (a), in fact, represents the chaotic attractor of system $(\ref{perioddoubling_original}).$  The extension of chaos through period-doubling cascade is observable in picture (b) such that the projection on the $x_3-x_4$ plane represents chaotic behavior. Appearance of the attractor in picture (b) around the limit cycle is a consequence of the entrainment process.}
\label{period_doubling_fig}
\end{figure}

Now, let us consider system $(\ref{perioddoubling_example1})$ with $\alpha=0.002$ and $\mu=0.008$ and take into account a trajectory of this system corresponding to initial data $x_1(0)=3.5, x_2(0)=-2,  x_3(0)=0.02$ and $x_4(0)=0.038.$ The $2-$dimensional projections of this trajectory on the $x_1-x_2$ and $x_3-x_4$ planes are presented in Figure \ref{period_doubling_fig}.  The picture in Figure $\ref{period_doubling_fig}, (a),$ illustrates the chaotic attractor of system $(\ref{perioddoubling_original})$ and the picture in Figure $\ref{period_doubling_fig}, (b),$ represents the chaos around limit cycle.  

Note that system $(\ref{perioddoubling_example1})$ exhibits a symmetry under the transformation 
\begin{eqnarray*}
\mathscr{K}:(x_1,x_2,x_3,x_4,t) \to (-x_1,-x_2,-x_3,-x_4,t+\pi).
\end{eqnarray*}
Therefore one can conclude that the chaotic attractor in the $4-$dimensional phase space of system $(\ref{perioddoubling_example1})$ is symmetric around the origin, and that is the reason for the symmetry of the projections on the $x_1-x_2$ and $x_3-x_4$ planes presented in Figure $\ref{period_doubling_fig}.$ 

In the next theorem, we will generalize our discussions about extension of chaos through period-doubling cascade in the case that system  $(\ref{2})$ is $n-$dimensional for arbitrary natural number $n.$ Before we state the theorem, we introduce the following definition.

We say that the solutions of the non-autonomous system 
\begin{eqnarray}
\upsilon'=h(t,\upsilon) \label{ultimately},
\end{eqnarray}
where $h:\mathbb R \times \mathbb R^q \to \mathbb R^q$ is a continuous function in all its arguments, are ultimately bounded for bound $B$ if there exists a number $B>0$ such that for every solution $\upsilon(t), \upsilon(t_0)=\upsilon_0,$ of system $(\ref{ultimately}),$ there exists a number $R>0$ such that the inequality $\left\| \upsilon(t) \right\|<B$ holds for all $t\geq t_0+R,$ where $B$ is independent of the particular solution while $R$ may depend on each solution. 

The proof of the next theorem can be verified using Theorem $15.8$ \cite{Yoshizawa75}.

\begin{theorem}\label{period-doubling_theorem}
If system $(\ref{1})$ admits the chaos through period-doubling cascade and if there exists a positive number $B$ such that for each chaotic solution $x(t)$ of system $(\ref{1})$ solutions of system $y'=f(y)+\mu g(x(t))$ are ultimately bounded for bound $B$, then system $(\ref{2})$ is chaotic in the same way.
\end{theorem}

We again emphasize that in the case of extension of chaos through period-doubling cascade, the unstability of the infinite number of periodic solutions of system $(\ref{2})$ is assured by Theorem $\ref{sensitivity_thm}.$ On the other hand, one can see that conditions of Theorem $(\ref{period-doubling_theorem})$ are fulfilled according to our discussions through Lyapunov functions for system $(\ref{perioddoubling_example1}),$  and this also provides existence of chaos through period-doubling cascade in the system.

\section{Discussion}

We shall devote this section to discuss through simulations the problems of entrainment of chaos by toroidal attractors, entrainment in Chua's oscillators and chaos control problem. We start with the demonstration of chaos generation around tori.

\subsection{\textbf{Entrainment of chaos by toroidal attractors}}

In previous parts of the paper, we have discussed entrainment of chaos with limit cycles. Now, the question is whether a similar approach is possible around tori. In this part, we will consider, numerically, the problem of seizure of chaos by tori.

Let us consider the following $3-$dimensional system  \cite{Hale91,Langford85}
\begin{eqnarray} \label{torus_eqn}
\begin{array}{l}
u_1'=(\lambda-3)u_1-0.25u_2+u_1\left(u_3+0.2(1-u_3^2)\right), \\
u_2'=0.25u_1+(\lambda-3)u_2+u_2\left(u_3+0.2(1-u_3^2)\right), \\
u_3'=\lambda u_3 -(u_1^2+u_2^2+u_3^2),
\end{array}
\end{eqnarray}
where $\lambda$ is a parameter.

For small and positive values of the parameter $\lambda,$ system $(\ref{torus_eqn})$ admits an asymptotically stable equilibrium point, with a positive $u_3$ coordinate near the origin, and at $\lambda \approx 1.68,$ the equilibrium point loses its stability and an hyperbolic, asymptotically orbitally stable limit cycle takes place. At the parameter value $\lambda=2,$ the periodic orbit is still asymptotically orbitally stable, but not hyperbolic. For $\lambda>2,$ the limit cycle is no longer stable and an attracting invariant torus takes place near the periodic orbit. With the increasing values of $\lambda,$ the invariant torus grows rapidly \cite{Hale91}.

\begin{figure}[ht] 
\centering
\includegraphics[width=10.5cm]{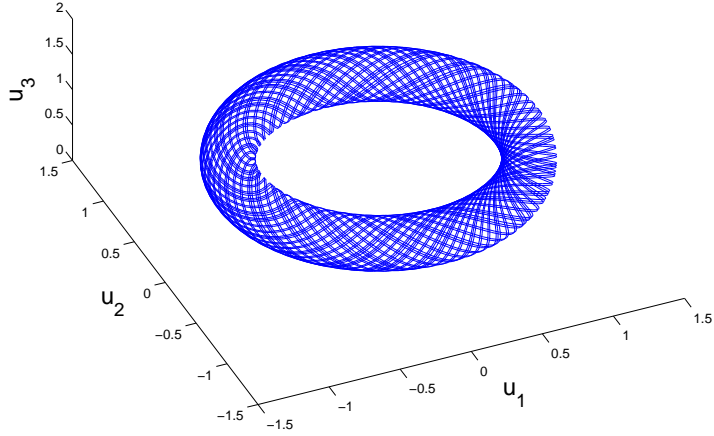}
\caption{The attracting invariant torus of system $(\ref{torus_eqn})$ with $\lambda=2.003.$}
\label{inv_torus}
\end{figure}

Figure $\ref{inv_torus}$ illustrates the trajectory of system $(\ref{torus_eqn})$ with the parameter value $\lambda=2.003,$ corresponding to the initial data $u_1(0)=-0.0983,  u_2(0)=0.9004$ and $u_3(0)=0.6908.$ It is seen in the figure that the motion of the trajectory is around a torus.

\begin{figure}[ht] 
\centering
\includegraphics[width=10.5cm]{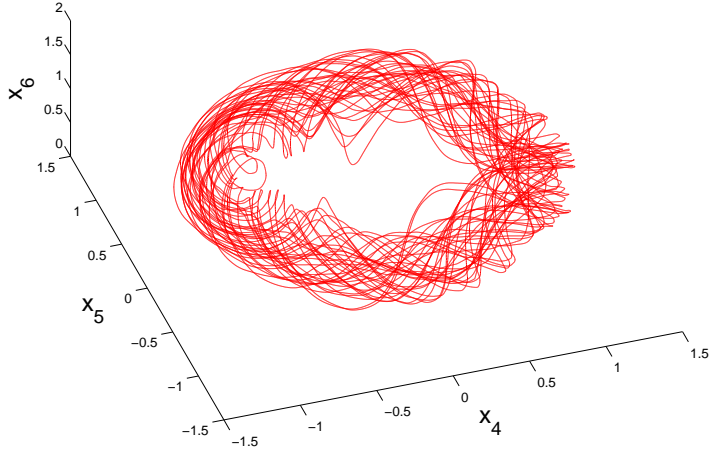}
\caption{The seizure of chaotic motion by  a  toroidal attractor. In the figure, the chaotic behavior around the invariant torus of system $(\ref{torus_eqn})$ with $\lambda=2.003$ is visualized. The picture is obtained by illustrating the $3-$dimensional projection of the chaotic trajectory of system $(\ref{torus_eqn2}),$ with initial data $x_1(0)=-6.7453,$ $x_2(0)=0.3435,$  $x_3(0)=32.7629,$   $x_4(0)=0.0793,$  $x_5(0)=-1.1761$ and  $x_6(0)=0.9449,$ on the $x_4-x_5-x_6$ space. The appearance of chaotic motion around a torus represents the entrainment process emerging in system $(\ref{torus_eqn2}).$ }
\label{chaotic_torus}
\end{figure}

To achieve chaotic motions around torus, we perturb system $(\ref{torus_eqn})$ through the solutions of the chaotic Lorenz system \cite{Sprott03}
\begin{eqnarray} \label{lorenz}
\begin{array}{l}
x_1'=-10x_1+10x_2,\\
x_2'=-x_1x_3+28x_1-x_2,\\
x_3'=x_1x_2-(8/3)x_3,
\end{array}
\end{eqnarray}
and constitute the following $6-$dimensional system
\begin{eqnarray} \label{torus_eqn2}
\begin{array}{l}
x_1'=-10x_1+10x_2,\\
x_2'=-x_1x_3+28x_1-x_2,\\
x_3'=x_1x_2-(8/3)x_3,\\
x_4'=(\lambda-3)x_4-0.25x_5+x_4\left(x_6+0.2(1-x_6^2)\right)\\+0.003x_1,\\
x_5'=0.25x_4+(\lambda-3)x_5+x_5\left(x_6+0.2(1-x_6^2)\right)\\+0.004x_2,\\
x_6'=\lambda x_6-(x_4^2+x_5^2+x_6^2)+0.002x_3,
\end{array}
\end{eqnarray}
where $\lambda=2.003$ once again.  Let us consider the solution of system $(\ref{torus_eqn2})$  corresponding to the initial data $x_1(0)=-6.7453,$ $x_2(0)=0.3435,$  $x_3(0)=32.7629,$  $x_4(0)=0.0793,$  $x_5(0)=-1.1761$ and  $x_6(0)=0.9449.$ Figure $\ref{chaotic_torus}$ shows the projection of the considered trajectory  on the $x_4-x_5-x_6$ space. It is observable in Figure $\ref{chaotic_torus}$ that the motion is disposed to behave both chaotically and around a torus, simultaneously.  

%The projection of the chaotic attractor of system $(\ref{torus_eqn2})$ on the $x_1-x_2-x_3$ space is the classical Lorenz attractor, which is not just pictured here. 

\subsection{\textbf{Entrainment in Chua's oscillators}}

We shall continue our investigation by presenting a simulation result for entrainment of chaos obtained in Chua's oscillators. The dimensionless form of a Chua's oscillator given by the system
\begin{eqnarray} 
\begin{array}{l}\label{Chua_1}
x_1'= k \alpha_0 [x_2-x_1 -\psi(x_1)], \\ 
x_2'=  k(x_1-x_2+x_3), \\
x_3'= k(-\beta_0 x_2-\gamma_0 x_3), \\
\psi(x)=bx+0.5(a-b)\left(  \left| x+1 \right|+ \left|x-1\right|  \right),
\end{array}
\end{eqnarray}
where $\alpha_0,\beta_0,\gamma_0, a, b$  and $k$ are constants.

In paper \cite{Chua93}, it is indicated that system $(\ref{Chua_1})$ with the coefficients $\alpha_0=\frac{21.32}{5.75}, \beta_0=7.8351, \gamma_0=\frac{1.38166392}{12}, a=-1.8459, b=-0.86604$ and $k=1$  admits a stable equilibrium.

We revealed in Section $\ref{sec_entrainment_period_doubling}$ that the system  $(\ref{perioddoubling_example1}),$ with the coefficients  $\alpha=0.002$ and $\mu=0.008,$ admits the chaos through period-doubling cascade and the $2-$dimensional projection on the $x_3-x_4$ plane of the chaotic attractor of this system appears near a limit cycle. Now, we attach two consecutive Chua's oscillators in dimensionless form, which admit asymptotically stable equilibrium points,    to constitute the following $10-$dimensional system 
\begin{eqnarray}
\begin{array}{l} \label{Chua_2}
x_1'=x_2, \\
x_2'=-0.3x_2-x_1^3+ 40, \displaystyle \cos t, \\
x_3'=0.002 x_3 -x_4 - x_3 \left(x_3^2+x_4^2\right) + 0.008 x_1, \\
x_4'= x_3 +0.002 x_4 - x_4 \left(x_3^2+x_4^2\right) + 0.008 x_2, \\
x_5'=\frac{21.32}{5.75} [x_6-0.13396x_5 +0.48993(  \left| x_5+1 \right|+ \left|x_5-1\right|  ) ] + 0.5x_3, \\ 
x_6'= x_5-x_6+x_7 + 2x_4, \\
x_7'=-7.8351 x_6-\frac{1.38166392}{12} x_7  + 3x_4, \\
x_8'=\frac{21.32}{5.75} [x_9-0.13396x_8 +0.48993 ( \left| x_8+1 \right|+ \left|x_8-1\right| ) ] +0.8x_5, \\ 
x_9'= x_8-x_9+x_{10} +0.3x_7, \\
x_{10}'=-7.8351 x_9-\frac{1.38166392}{12} x_{10} +0.8x_6.
\end{array}
\end{eqnarray}

Since the subsystem with coordinates $(x_1,x_2)$ has chaos through period-doubling cascade, the subsystem $(x_3,x_4)$ admits the seized chaos by its limit cycle, which is guaranteed by our theoretical discussions. To illustrate an application of our results, we make use of system $(x_3,x_4)$ as a source of chaos for the Chua circuit, presented by the subsystem $(x_5,x_6,x_7).$  In addition,  the coordinates $x_5, x_6$ and $x_7$ are used to perturb the next Chua's oscillator corresponding to the last three equations in system $(\ref{Chua_2}).$  Following the results of paper  \cite{Akh8}, we have to observe chaotic behavior in both of the Chua's oscillators.

\begin{figure}[ht] 
\centering
\includegraphics[width=10.5cm]{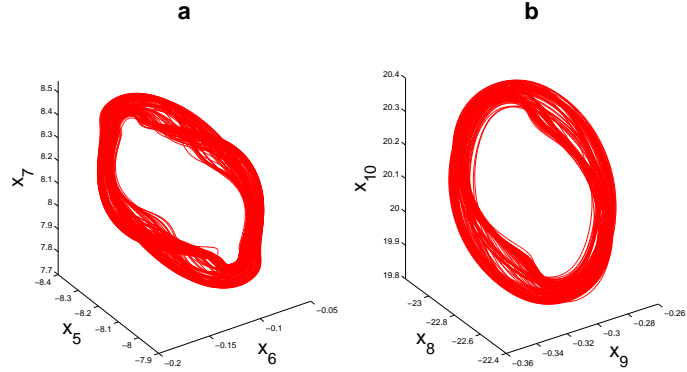}
\caption{$3-$dimensional projections of the chaotic attractor of system $(\ref{Chua_2}).$ (a) Projection on the $x_5-x_6-x_7$ space, (b) Projection on the $x_8-x_9-x_{10}$ space.  The pictures presented in $(a)$ and $(b)$ both reveal that the Chua's oscillators possess chaotic behavior such that the chaotic attractors appear near limit cycles, and this is an indicator of entrainment of chaos.}
\label{chua_fig}
\end{figure}

We consider a trajectory of system $(\ref{Chua_2})$ with initial data $x_1(0)=3.5,$  $x_2(0)=-2,$  $x_3(0)=0.02,$  $x_4(0)=0.038,$  $x_5(0)=-8.016,$  $x_6(0)=-0.084,$ $x_7(0)=7.792,$ $x_8(0)=-22.764,$ $x_9(0)=-0.281$ and $x_{10}(0)=20.167,$ and visualize its $3-$dimensional projections on the $x_5-x_6-x_7$ and $x_8-x_9-x_{10}$ spaces in Figure $\ref{chua_fig}.$ We note that the projections on the $x_1-x_2$ and $x_3-x_4$ planes will give the same attractors presented in Figure $\ref{period_doubling_fig}, (a)$ and $(b),$ respectively. The pictures presented in Figure $\ref{chua_fig}, (a)$ and $(b)$ indicate that the chaotic Chua's attractors appear around limit cycles and this is a manifestation of entrainment of chaos.  Furthermore, it is seen in Figure $\ref{chua_fig}$ that the shapes of these attractors resemble the spiral Chua's attractor, which takes place in the case of a period-doubling cascade in Chua systems \cite{Chua93,Lakshmanan03}.

\subsection{\textbf{Controlling chaos}}

The chaos obtained through period-doubling cascade admits infinitely many periodic solutions which are all unstable and Pyragas control method \cite{Pyragas92} is an effective instrument to stabilize them. One can say that it plays also an important role as it is very convenient to indicate through simulations the periodic solutions, which are invisible in the set of irregular motions.

In this subsection, we will indicate a procedure by favour of an example for the stabilization of unstable periodic solutions of system $(\ref{1})+(\ref{2}).$ The Pyragas control method will be mainly used in this procedure and therefore we start by explaining the method briefly.  Pyragas, in 1992, proposed a delayed feedback control method for the stabilization of unstable periodic orbits of a chaotic system. In this method, one considers a system of the form
\begin{eqnarray}
\begin{array}{l}
x'=H(x,q), \label{pyragas_controll} 
\end{array}
\end{eqnarray}
where $q=q(t)$ is an externally controllable parameter and for $q=0$ it is assumed that the system $(\ref{pyragas_controll})$ is in the chaotic state of interest, whose periodic orbits are to be stabilized \cite{Zelinka,Pyragas92,Gon04,Sch08,Fra07}. According to Pyragas method, an unstable periodic solution with period $\tau_0$ of system $(\ref{pyragas_controll})$ with $q=0,$ can be stabilized by the control law $q(t)=C \left[ s\left(t- \tau_0 \right) -s(t) \right],$ where the parameter $C$ represents the strength of the perturbation and $s(t)= \sigma \left[x(t)\right]$ is a scalar signal given by some function of the state of the system.

It is indicated in \cite{Gon04} that in order to apply the Pyragas control method to the chaotic Duffing oscillator given by the system
\begin{eqnarray}
\begin{array}{l}
x_1'=x_2, \\  \label{pyragas_control2}
x_2'=-0.10 x_2 + 0.5 x_1 \left( 1-x_1^2  \right) + 0.24 \displaystyle \sin t, \\
\end{array}
\end{eqnarray}
one can  construct the corresponding control system
\begin{eqnarray}
\begin{array}{l}
z_1'=z_2, \\  \label{pyragas_control3}
z_2'=-0.10 z_2 + 0.5 z_1 \left( 1-z_1^2  \right) \\
+ 0.24 \displaystyle \sin(z_3) + C \left[ z_2(t-\tau_0) -z_2(t)  \right], \\
z_3'=1,
\end{array}
\end{eqnarray}
where $q(t)=C \left[ z_2(t-\tau_0) -z_2(t)  \right]$ is the control law and the less unstable $2\pi-$periodic solution can be stabilized by choosing the parameter values $C=0.36$ and $\tau_0=2\pi.$

Making use of system $(\ref{pyragas_control2})$ together with $(\ref{orbitally_stable_system}),$ where $\alpha=7,$ we set up the following system
\begin{eqnarray}
\begin{array}{l} \label{control_1}
x'_1=x_2, \\
x'_2=-0.10x_2+0.5x_1(1-x_1^2)+0.24\displaystyle \sin t, \\
x'_3=7x_3-x_4-x_3(x_3^2+x_4^2)+5x_1, \\
x'_4=x_3+7x_4-x_4(x_3^2+x_4^2)+4(x_2+x_2^3).
\end{array}
\end{eqnarray}

According to our theoretical discussions, system $(\ref{control_1})$ admits a  chaotic attractor in the $4-$dimensional phase space, and its $2-$dimensional projection on the $x_3-x_4$ plane appears near the limit cycle of system $(\ref{orbitally_stable_system}),$ with $\alpha=7.$

Our present purpose is to demonstrate numerically how to  control chaos of system $(\ref{control_1}).$ We propose that if a periodic solution of the $2-$dimensional subsystem $(x_1,x_2),$ inside system $(\ref{control_1}),$ is stabilized, then the chaos of system $(\ref{control_1})$ is controlled. In other words, it is enough to control the chaos of system $(\ref{pyragas_control2}),$ which is used as the source of the exogenous perturbation in system  $(\ref{control_1}).$

To apply the Pyragas  method for controlling the chaos of system $(\ref{control_1}),$ we constitute the system
\begin{eqnarray}
\begin{array}{l} \label{control_2}
z'_1=z_2, \\
z'_2=-0.10z_2+0.5z_1(1-z_1^2)+0.24\displaystyle \sin z_3+C \left[ z_2(t-2\pi) -z_2(t)  \right], \\
z'_3=1, \\
z'_4=7z_4-z_5-z_4(z_4^2+z_5^2)+5z_1, \\
z'_5=z_4+7z_5-z_5(z_4^2+z_5^2)+4(z_2+z_2^3),
\end{array}
\end{eqnarray}
which is the control system corresponding to system $(\ref{control_1}).$ 

\begin{figure}[ht] 
\centering
\includegraphics[width=12.5cm]{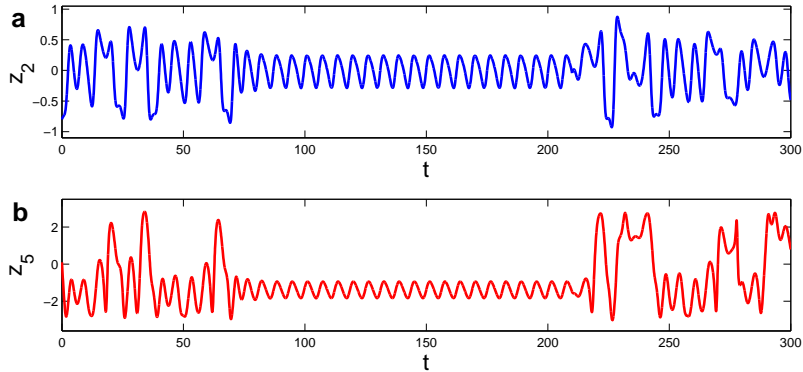}
\caption{Application of the Pyragas control method to system $(\ref{control_1})$ by means of the corresponding control system $(\ref{control_2}).$ (a) The graph of $z_2$ coordinate, (b) The graph of $z_5$ coordinate. The picture in (a) reveals that the control mechanism stabilizes the less unstable $2\pi-$periodic solution of system $(\ref{pyragas_control2}),$ and correspondingly the picture in (b) supports that the chaos of system $(\ref{control_1})$ is controlled. The control mechanism starts at the time $t=70$ and ends at $t=210.$ On the intervals where the control is switched off, the irregular behavior of the coordinates are observable.}
\label{control_fig}
\end{figure}

We consider the solution of system $(\ref{control_2})$ with initial data $z_1(0)=0.1,z_2(0)=-0.8,z_3(0)=0,z_4(0)=2.64$ and $z_5(0)=0.1.$ The system evolves freely taking $C=0$ until $t=70,$ and at that moment the control is switched on by taking $C=0.36.$ At the moment $t=210,$ we switch off the control mechanism and start to use the value of the parameter $C=0$ once again. Figure $\ref{control_fig}$ pictures the graphs of the $z_2$ and $z_5$ coordinates of the solution which reveals the control of chaos of system $(\ref{control_1}).$ It is also observable that after switching off the control mechanism, the stabilized $2\pi-$periodic solution of system $(\ref{control_1})$ loses its stability and chaos emerges again. Similar pictures can be obtained for the other coordinates of system $(\ref{control_2}),$ which are not just pictured here.

\section{Conclusion} \label{conclusion}

The ideas of Huygens about synchronization are carried a step forward in such a way that the entrainment of chaos by limit cycles is discussed. Our theoretical results can be effectively used in arbitrarily high dimensional systems, which possess  asymptotically orbitally stable limit cycles. Examples of such systems can be found in mechanics, electronics, economics, neural sciences, chemistry and population dynamics \cite{Lorenz93,Hassard81,Field93,Kostova04,Jiang04,Wang08,Morton99}. Through the method presented, one can obtain motions which behave cyclically and chaotically in the same time. 

In the paper, theoretical problems with rigorous proofs for the existence of bounded solutions and extension of sensitivity, which we consider as the unique ingredient of chaos, as well as extension of chaos obtained through period-doubling cascade are considered. Additionally, the illustrated simulations support the theoretical results. Entrainment of chaos by toroidal attractors and entrainment in Chua's oscillators have been observed numerically. We indicate the existence of a chaotic attractor in a sample system by means of Lyapunov functions. Moreover, the existence of unstable periodic solutions is discussed  through the Pyragas method \cite{Pyragas92} and simulations.

Cyclical behavior in chaotic attractors have been widely observed in the literature. We can refer for this famous R$\ddot{o}$ssler attractor and Chua's spiral attractor, and even in the classical Lorenz attractor one can see two-center cyclical behavior. Our results for the achievement of cyclical behavior  are different than those, since ``cycling" chaos is usually obtained through period-doubling cascade, while our irregular cyclic behavior is a result of already existing chaotic motions, which are not in general ``cyclic", but applied as exogeneous perturbations. Nevertheless, one can guess that the mechanism proposed in our article may be underneaths of some chaotic attractors, which have already been discussed in the literature.  

The results of the paper can be mimicked for the case when cycles are attracting for the time decreasing to $-\infty.$ Moreover, they can be extended by considering tori as attractors. Another theoretically challenging problem is to consider hyperbolic cycles and also the critical cases.

Our results are useful to generate multidimensional chaos, exceptionally if one requests that the phenomenon should be rigorously approved \cite{Marotto78,Marotto05}. If one considers Hopf bifurcation as a reason for the limit cycle generation, we can formally compare our results with the results of Ruelle and Takens \cite{Ruelle71} on the appearance of  turbulence through three successive bifurcations. In our case, we have a chaos obtained after less than three bifurcations and additionally, we use chaotic perturbations.

\vspace{0.55cm}
\noindent{\bf{\Large Acknowledgements}}

This research was supported by a grant (111T320) from TUBITAK, the Scientific and Technological Research Council of Turkey.

%\noindent\textbf{References}

\end{document}